\documentclass[traditabstract]{aa} 
\usepackage{graphicx}
\usepackage{txfonts}
\usepackage{natbib}
\bibpunct{(}{)}{;}{a}{,}

\newcommand\lesim{\lower.5ex\hbox{$\; \buildrel < \over \sim \;$}}
\newcommand\gesim{\lower.5ex\hbox{$\; \buildrel > \over \sim \;$}}
\newcommand\HI{H{\small I}}
\newcommand\kms{km s$^{-1}$}
\newcommand\Ha{H$\alpha $}
\newcommand\mjybeam{mJy beam$^{-1}$}
\newcommand\atomscmsq{atoms cm$^{-2}$}

\usepackage{color}
\definecolor{red}           {cmyk}{0.   , 1.   , 1.   , 0.   }
\definecolor{lightred}      {rgb}{1.00, 0.35, 0.05}
\definecolor{darkred}       {rgb}{0.70, 0.05, 0.05}
\definecolor{lightblue}      {rgb}{0.05, 0.35, 1.00}
\definecolor{darkblue}       {rgb}{0.05, 0.05, 0.65}
\definecolor{lightpurple}    {rgb}{1.00, 0.00, 1.00}
\definecolor{darkpurple}     {rgb}{0.65, 0.00, 0.65}
\definecolor{lightgreen}     {rgb}{0.05, 0.90, 0.05}
\definecolor{darkgreen}       {rgb}{0.05, 0.65, 0.05}
\definecolor{brown}           {rgb}{0.80, 0.60, 0.05}
\definecolor{green}           {cmyk}{ 1. ,0.  , 1., 0.   }
\definecolor{yellow}           {cmyk}{ 0., 0.  , 1., 0.   }
\definecolor{maroon}           {cmyk}{ 0.16 ,0.98  , 1., 0.13   }
\definecolor{black}           {cmyk}{ 0. ,0.  , 0., 0.   }

\begin{document}
   \title{The dark matter halo shape of edge-on disk galaxies}

\subtitle{I. HI observations}

\titlerunning{The dark matter halo shape of edge-on disk galaxies I}
   \author{J.C. O'Brien
          \inst{1},
K.C. Freeman\inst{1},
          P.C. van der Kruit\inst{2}
\and
A. Bosma\inst{3}}
   \institute{Research School of Astronomy and Astrophysics, 
Australian National
University, Mount Stromlo Observatory, Cotter Road, ACT 2611, Australia\\
              \email{jesscobrien@gmail.com; kcf@mso.anu.edu.au$^\star$}
\and     
             Kapteyn Astronomical Institute, University of Groningen, 
P.O. Box 800,
9700 AV Groningen, the Netherlands\\
             \email{vdkruit@astro.rug.nl}\thanks{For 
correspondence contact Ken Freeman or 
Piet van der Kruit}
\and
 Laboratoire d'Astrophysique de Marseille, UMR 6110 Universite 
de Provence / CNRS, 38 rue Fr\'ed\'eric Joliot-Curie, 13388 Marseille
C\'edex 13, France\\
\email{bosma@oamp.fr}
             }
\authorrunning{J.C. O'Brien et al.}

   \date{Received Xxxxxxx 00, 2010; accepted Xxxxxxx 00, 2010}

 
  \abstract
   {This is the first paper of a series in which we will attempt to put
constraints on the flattening of dark halos in disk galaxies. We
observe for this purpose the \HI\ in edge-on galaxies, where it is
in principle possible to measure the force field in the halo
vertically and radially from gas layer flaring and rotation curve
decomposition respectively.  In this paper, we define a sample
of 8 \HI\ rich late-type galaxies suitable for this purpose and
present the \HI\ observations.
%
   \keywords{galaxies: structure; galaxies: 
kinematics and dynamics; galaxies: halos; galaxies: ISM}
}
   \maketitle
%

\section{Introduction}

Since the 1970's, it has been known that the curvature of the universe
is remarkably flat. This implies that the ratio of the total density
to critical density is $\Omega_{tot} \approx 1$.  At this time, it
was also known from measurements of stars and gas that the mass
density of luminous matter is $\Omega_{lum} \lesim 0.005$ --- less
than $0.5\%$ of that required for a flat $\Omega_{tot}\sim$1 universe.
Indeed, the missing matter controversy had begun in the
1930's when \citet{oort1932}, following \citet{kapteyn1922}, and \citet{zwicky1937} 
independently found
evidence for vast amounts of unseen matter on different scales.
Oort's analysis of the spatial and velocity distribution of stars in the Solar
neighbourhood concluded that luminous stars comprised approximately 
half the total mass indicated by their motion, assuming
gravititational equilibrium.  Zwicky's analysis of the velocity dispersions
of rich clusters found that approximately $90-99\%$ of the mass was unseen, 
if the systems were gravitationally bound. Later on, the
inventory of the mass in the Solar neighbourhood showed that the dark
matter in the galactic disk at the solar galactocentric radius was mostly fainter
stars and interstellar gas \citep[e.g.][]{hf00}. Zwicky's
value of the velocity dispersion in the Coma cluster is close to the
current one \citep[e.g.][]{cd96}, and, despite the presence of hot X-ray 
emitting gas, there must be dark matter to cause $\Omega_{matter} 
\sim 0.2 - 0.3$.

\begin{table*}[t] 
\centering
\caption[HI observations]{HI observations}
\label{tab:obs-HI}
\begin{small}
\begin{tabular}{lllccllc}
\hline
Galaxy & Other & Date & Telescope & Array & Project & Observer
& Integration \\
& name &&&& \ \ ID && time (hr) \\
\hline\hline
ESO074-G015 & IC5052  & 11-12 FEB 2001 & ATCA & 375   & C934    & Ryan       &     1.67 \\
ESO074-G015 & IC5052  & 13-14 APR 2001 & ATCA & 750D  & C934    & Ryan       &     2.21  \\
ESO074-G015 & IC5052  & 24-25 FEB 2002 & ATCA & 1.5A  & C894    & O'Brien    &    10.23   \\
ESO074-G015 & IC5052  & 01 DEC 2002    & ATCA & 6.0A  & C894    & O'Brien    &    10.63   \\
ESO074-G015 & IC5052  & 17 OCT 1992    & ATCA & 6.0C  & C212    & Carignan   &    10.05   \\
\hline
ESO109-G021 & IC5249  & 20-22 MAR 2003 & ATCA & EW352 & CX043   & Dahlem     &     8.54   \\
ESO109-G021 & IC5249  & 03-04 FEB 2003 & ATCA & 750D  & C894    & O'Brien    &     8.01  \\
ESO109-G021 & IC5249  & 28 NOV 2002    & ATCA & 6.0A  & C894    & O'Brien    &    13.43   \\
ESO109-G021 & IC5249  & 18 OCT 1992    & ATCA & 6.0C  & C212    & Carignan   &    11.11  \\
\hline
ESO115-G021 &         & 09 FEB 2005    & ATCA & EW352 & C1341   & Koribalski &    10.09   \\
ESO115-G021 &         & 06 JAN 2005    & ATCA & 750B  & C1012   & Hoegaarden &    11.17   \\
ESO115-G021 &         & 23 JUN 1995    & ATCA & 750B  & C073    & Walsh      &     7.57   \\
ESO115-G021 &         & 08 SEP 2002    & ATCA & 6.0C  & C894    & O'Brien    &     6.51  \\
ESO115-G021 &         & 03 DEC 2002    & ATCA & 6.0A  & C894    & O'Brien    &    10.05   \\
ESO115-G021 &         & 13 DEC 2002    & ATCA & 6.0A  & C894    & O'Brien    &     5.72  \\
\hline
ESO138-G014 &         & 08-09 NOV 2002 & ATCA & 1.5A  & C894    & O'Brien    &    10.43   \\
ESO138-G014 &         & 29-30 NOV 2002 & ATCA & 6.0A  & C894    & O'Brien    &     9.79   \\
\hline
ESO146-G014 &         & 17 JAN 2002    & ATCA & 750A  & C894    & O'Brien    &     1.37   \\
ESO146-G014 &         & 27-29 DEC 2000 & ATCA & 750C  & C894    & O'Brien    &    10.70  \\
ESO146-G014 &         & 11 JAN 2001    & ATCA & 750C  & C894    & O'Brien    &     1.67  \\
ESO146-G014 &         & 31 JUL 2001    & ATCA & 1.5A  & C894    & O'Brien    &     7.78  \\
ESO146-G014 &         & 14-15 APR 2002 & ATCA & 6.0A  & C894    & O'Brien    &    10.19 \\
ESO146-G014 &         & 27-28 JAN 2002 & ATCA & 6.0B  & C894    & O'Brien    &    10.51  \\
\hline
ESO274-G001 &         & 29 AUG 1993    & ATCA & 1.5B  & C073    & Walsh      &     6.07  \\
ESO274-G001 &         & 07 OCT 1993    & ATCA & 1.5D  & C073    & Walsh      &     6.80   \\
ESO274-G001 &         & 28-29 NOV 2002 & ATCA & 6.0A  & C894    & O'Brien    &     9.63   \\
\hline
ESO435-G025 & IC2531  & 12 JAN 2002    & ATCA & 750A  & C894    & O'Brien    &     9.63  \\
ESO435-G025 & IC2531  & 17 JAN 2002    & ATCA & 750A  & C894    & O'Brien    &     2.02  \\
ESO435-G025 & IC2531  & 07 MAR 1997    & ATCA & 1.5D  & C529    & Bureau     &     9.07  \\
ESO435-G025 & IC2531  & 06 APR 1996    & ATCA & 6.0A  & C529    & Bureau     &     9.94  \\
ESO435-G025 & IC2531  & 13-14 SEP 1996 & ATCA & 6.0B  & C529    & Bureau     &     9.96 \\
ESO435-G025 & IC2531  & 17-18 OCT 1992 & ATCA & 6.0C  & C212    & Carignan   &     7.45 \\
\hline
UGC07321    &         & 26,30 MAY 2000 & VLA  & C     & AM649   & Matthews   &    16.00   \\
UGC07321    &         & 01 NOV 2000    & ATCA & 1.5D  & C894    & O'Brien    &     2.49  \\
\hline
\end{tabular}
\end{small}
\end{table*}

Modern dark matter research began in 1970 with several papers
which found that galaxies contain more gravitating matter than can
be accounted for by the stars.  \citet{freeman1970} noted
that the atomic hydrogen (\HI) rotation curves of the late-type disk
galaxies NGC300 and M33 peaked at a larger radius than expected
from the stellar light distribution. This implied that the dark
matter was more extended than the stellar distribution and that its
mass was at least as great as the luminous mass. 
However, these data were of poor angular
resolution. At the same time \citet{rubin1970} published a rotation
curve of M31 based on optical data, which did not seem to decline in 
the outer parts.
In the 1970s radio observations of increasingly better angular
resolution and better sensitivity\footnote{KCF and PCvdK recall 
influential colloquia and other presentations by M.S. Roberts 
in the early 1970s on a flat \HI\ rotation curve of the Andromeda galaxy
which helped to steer the evolution of this subject.}  
 showed that flat rotation curves
are typical in spiral galaxies \citep{shostak1971,rs1972,rr1973,bosma1978}
and that the HI extent of a spiral galaxy can be far greater than
the extent of the optical image. This, combined with surface
photometry, leads to very high mass-to-light ratios in the outer 
parts, as shown clearly for several galaxies in \citet{bosma1978} and
\citet{bk1979}.

In 1974, dark
matter halos were found to extend even further, when \citet{oyp1974}
and \citet{eks1974} tabulated galaxy masses as a function of radius
and found that galaxy mass increased linearly out
to at least $100$ kpc and $10^{12}$ M$_{\odot }$ for normal spirals
and ellipticals.  Despite this large dark matter fraction inferred
in galaxies, it was still not sufficient to reach the critical
$\Omega_{tot}\sim$1 value of a flat universe.  See e.g. 
\citet{sr01} and \citet{rob2008}
for a brief history of dark matter in galaxies.

Around the same time, application of nuclear physics to Big Bang
theory showed that big bang nucleosynthesis (BBNS) in the early 
universe produced specific abundances of the light
elements, and predicted the total baryon density was $\Omega_b
\approx 0.044$.  
In the last decade, the combined observations of high redshift SNIa,
\citep{reissetal1998,perlmutteretal1999}, the 2-Degree Field Galaxy
Redshift Survey (2dFGRS) \citep{percivaletal2001} and WMAP microwave 
background measurements \citep{spergel_wmap2003} confirmed the baryonic 
mass density $\Omega_b$ found by BBNS and established
that dark energy comprises about $75\%$ of the critical density.  
Consequently, the scale of the missing
dark matter is now reduced. The mass density $\Omega_m$ is now only
$0.25$, but the problem remains: the measured baryonic mass density
is still only $\sim$18\%\ of the total mass density $\Omega_m$.  
Thus, dark matter accounts for $82\%$ of mass in the universe, and 
baryonic matter is only 4.5\%\ of the total content of the 
universe.\footnote{For this illustration we used the
parameters adopted in the Millenium Simulation \citep{swjetal2005}, 
which are $\Omega_m = \Omega_{dm} + \Omega_b = 0.25$, $\Omega_b = 0.045$,
$\Omega_{\Lambda} = 0.75$.} 

More than 1000 galactic rotation curves have now been measured, and
very few display a Keplerian decline with radius. \HI\ and \Ha\
rotation curves of spiral galaxies show that the total-mass-to-light
ratios are typically M/L = 10-20 M$_{\odot}$/L$_{\odot}$, and the
luminous matter therefore accounts for only $5-10\%$ of the total
mass inferred from the rotation curves.  For low surface brightness
(LSB) disk galaxies \citep{deblok1997} and dwarf irregular (dI)
galaxies \citep{swaters1999}, the M/L values increase to 10-100
M$_{\odot}$/L$_{\odot}$, with an extreme of $220$ for ESO215-G009,
a gas-rich LSB galaxy with a very high gas mass to light ratio of
$M_{\rm HI}/L_B = 21$ and low recent star formation \citep{wjk2004}.

Dwarf spheroidal (dSph) galaxies, with typical total masses of only
$\sim$10$^{7}$ M$_{\odot }$ within radii of a few hundred parsecs,
are even more extreme: several have very large dark matter fractions
with mass-to-light ratios in the range $200-1000$ M$_{\odot}$/L$_{\odot}$.
In these faint, small galaxies, the dynamical mass is estimated
from the line-of-sight velocities of individual stars.  The Ursa
Major dSph \citep{willmanetal2005}, is one of the most dark matter
dominated galaxies known to date with a central mass-to-light ratio
$M/L\sim$500 M$_{\odot}$/L$_{\odot}$, which is believed to increase
further at larger radii \citep{kweg2005}.  These systems appear to
have only very small baryonic mass fractions.

The rotation curves of disk galaxies are important probes
of the equatorial halo potential gradient. By decomposing the 
observed rotation curve into contributions from the visible mass 
components, the radial potential gradient of the
halo can be measured, assuming the system is in centrifugal 
equilibrium. The dark halo mass component is typically fitted
by a pseudo-isothermal halo model with density 
\begin{equation}
\label{eq:pISO} 
\rho(R) = \rho_0 \left[ 1 + \left(\frac{R}{R_c}\right)^2 \right]^{-1},
\end{equation}  
where the halo is characterised by its
central density $\rho_0$ and core radius $R_c$. Pseudo-isothermal halos have
an asymptotic density $\rho \propto R^{-2}$ at large radii 
which is consistent with commonly flat rotation curves.

\citet{kf2004} compiled the published dark halo density distributions
for a large sample of Sc-Im and dSph galaxies, and found well-defined
scaling relationships for the dark halo parameters for galaxies
spanning a range of 6 decades of luminosity.\footnote{E-Sbc galaxies
were not included to avoid the larger uncertainties associated with
stellar bulge-disk decomposition and the relatively larger contribution
of the stellar mass of varying stellar ages.}  They found that halos
of less luminous and massive dwarf spheroidals have higher central densities,
up to $\sim$1 M$_{\odot}$ pc$^{-3}$  and
core radii of $\sim$0.1 kpc, compared to $\sim10^{-3}$ M$_{\odot}$
pc$^{-3}$ and $\sim$30-100 kpc respectively, for large bright Sc
galaxies. The observed correlations suggest a continuous physical
sequence of dark halo population in which the properties of the
underlying dark halo scale with the baryon luminosity \citep{kf2004}.

We now consider the flattening of the dark halo density distribution,
defined by $q=c/a$, where $c$ is the halo polar axis and $a$ is the
major axis in the galactic plane.  The vertical distribution of the
halo is much more difficult to measure than the radial distribution
in the equatorial plane, as most luminous tracers of the galaxy
potential gradient lie in the plane of the galaxy and offer little 
indication of the vertical gradient of the potential.  Past measurements 
that were obtained
with a variety of different methods gave a large range of $q$ from $0.1$ to
$1.37$, with no concentration on any particular value.  For
our Galaxy, the halo shape has been measured more than ten times
by four different methods, that yielded $q-$values ranging from $0.45$ to
$1.37$.

In this series of papers, we will use the flaring
of the HI gas layer to measure the vertical flattening of the dark halo, 
because we believe this method to be the most promising for
late-type disk galaxies.  Like all steady state mass components
of a galaxy, the gaseous disk of an isolated disk galaxy can be
assumed to be in hydrostatic equilibrium in the gravitational
potential of the galaxy, unless there are visible signs from the
gas distribution and kinematics that the HI layer is disturbed
(e.g. by mergers or local starbursts). In the vertical direction,
the gradient of gas pressure is balanced by the 
gravitation force (ignoring any contribution from
a magnetic pressure gradient). From the observed distribution of the gas velocity
dispersion and the gas density distribution, we can in principle
measure the total vertical gravitational force.

\begin{table}[t] 
\centering 
\caption[Spatial \& spectral resolution of \HI\ observations]{Resolution of 
\HI\ observations}
\label{tab:resol-HI}
\begin{tabular}{lccc}
\hline
 & \multicolumn{2}{c}{Spatial} & Spectral \\
Galaxy & \multicolumn{2}{c}{resolution} & resolution \\
\cline{2-3}
& (arcsec) & (kpc) & (km s$^{-1}$)\\
\hline\hline
ESO074-G015 &   9.0 $\times$ 9.0  & 0.292 $\times$ 0.292 &   3.298 \\
ESO109-G021 &   8.0 $\times$ 8.0  & 1.179 $\times$ 1.179 &   3.298 \\
ESO115-G021 &   8.9 $\times$ 8.9  & 0.194 $\times$ 0.194 &   3.298 \\
ESO138-G014 &  10.7 $\times$ 10.7 & 0.965 $\times$ 0.965 &   3.298 \\
ESO146-G014 &   7.6 $\times$ 7.6  & 0.793 $\times$ 0.793 &   3.298 \\
ESO274-G001 &   9.8 $\times$ 9.8  & 0.162 $\times$ 0.162 &   3.298 \\
ESO435-G025 &   9.0 $\times$ 9.0  & 1.305 $\times$ 1.305 &   3.298 \\
UGC07321    &  15.0 $\times$ 15.0 & 0.727 $\times$ 0.727 &   5.152 \\
\hline
\end{tabular}
\end{table}

Euler's equation for a steady-state fluid of density $\rho$, velocity
$\mathbf{v}$ and pressure $p$ in a gravitational potential $\Phi$ 
is $-\mathbf{(v.\nabla)v} = \rho^{-1}
\mathbf{\nabla}p + \mathbf{\nabla}\Phi$.  
In the case of a vertically Gaussian gas
density distribution with a vertically isothermal gas velocity
dispersion, the gradient of the total vertical force $K_z$ in the
$z$-direction can be calculated directly from the gas velocity
dispersion $\sigma_{v,g}(R)$ and the gas layer thickness FWHM$_{z,g}(R)$,
each of which are functions of radius:
\begin{equation}
\label{eq:ch1-hydro}
\frac{\partial K_z}{\partial z} = -
\frac{\sigma_{v,g}^2}{({\rm FWHM}_{z,g}/2.35)^2}.
\end{equation}
It is possible to measure the
halo shape over the entire \HI\ extent of the luminous disk using the 
flaring of the \HI\ distribution, which typically extends
in radius to $2-3 R_{25}$, and by measuring the density distribution
of the gas and stellar distributions.  For
a given vertical gas velocity dispersion, a more flattened
dark halo requires decreased flaring
and increased gas surface density. The relatively high gas content of 
late-type galaxies allows measurement of both the halo vertical force 
field from the gas layer 
flaring and the halo radial force field from rotation curve decomposition.
In this series of papers we will attempt to measure the halo flattening
using the flaring of \HI\ disk in eight small late-type disk galaxies.

\begin{table}[t] 
\caption[Noise of \HI\ channel maps]{Noise of \HI\ channel maps}
\label{tab:noise-HI}
\begin{tabular}{lcccc}
\hline
Galaxy & \multicolumn{3}{c}{Noise} & Max Signal-\\
\cline{2-4}
& (Jy/beam) & (atoms cm$^{-2}$) & (K) & to-noise \\
\hline\hline
ESO074-G015 &  0.00135 &    1.8348 $10^{19}$ &   10.0 & 14.5 \\
ESO109-G021 &  0.00125 &    2.1464 $10^{19}$ &   11.7 & 10.1 \\
ESO115-G021 &  0.00113 &    1.5774 $10^{19}$ &    8.6 & 18.6 \\
ESO138-G014 &  0.00197 &    1.9008 $10^{19}$ &   10.4 & 15.2 \\
ESO146-G014 &  0.00143 &    2.7190 $10^{19}$ &   14.9 & 11.3 \\
ESO274-G001 &  0.00151 &    1.7209 $10^{19}$ &    9.4 & 14.6 \\
ESO435-G025 &  0.00123 &    1.6789 $10^{19}$ &    9.2 & 10.9 \\
UGC07321    &  0.00038 &    1.8579 $10^{18}$ &    1.0 & 93.0 \\
\hline
\end{tabular}
\end{table}

The main advantage of this method is that it can be used for all
gas-rich spiral galaxies inclined sufficiently to measure accurate
kinematics, unlike some methods which are applicable only to specific 
kinds of galaxies like polar ring galaxies. 
This minimum inclination was determined to be 
$i \gesim 60^{\circ}$ by \citet{olling1995}.
This method was first tried by \citet{crb1979} on the Galaxy, 
and early development was undertaken by \citet{vdkruit1981}, who 
applied it to low resolution observations of NGC891 and
concluded that the halo was
not as flattened as the stellar disk. It was then used for several
galaxies in the 1990's, most notably the careful study of the very
nearby Sc galaxy NGC4244 which was found to have a highly flattened 
halo with 
$q=0.2^{+0.3}_{-0.1}$ out to radii of $\approx 2 R_{25}$
\citep{olling1996b}.

All applications of the \HI\ flaring method to date have derived
highly flattened halo distributions ($q \leq 0.5$). With the exclusion
of NGC4244, we suspect that the assumption of a radially-constant
gas velocity dispersion led to errors in the measured total vertical
force, and thereby to the derived flattening of
the halo. The other difficulty for all these galaxies except
NGC4244 is that they are large galaxies with peak rotation speeds
$>170$ \kms. Given that {\it (i)} spiral galaxies typically have
\HI\ velocity dispersions in the relatively small range $6-10$ \kms 
\citep{tamburro09}
and {\it (ii)} the halo shape $q$ is roughly proportional
to the gradient of the vertical force, we see from
Eqn.~\ref{eq:ch1-hydro} that the \HI\ thickness is to first order
inversely proportional to the peak rotation speed. Consequently, 
the \HI\ flaring
method should work best for small disk galaxies with relatively low
rotation speeds, as these galaxies should exhibit more \HI\ flaring.
\citet{bosma1994} already showed by
calculations that flaring is relatively more important in galaxies
with low circular velocity.

The flaring method has also been applied to the Galaxy by \citet{om2000}.
However, uncertainty in the position and rotation speed of the Sun
resulted in a large uncertainty of the measured halo shape $0.5
\leq q \leq 1.25$.

\begin{table*}[t] 
\centering
\caption[HI measurements]{HI measurements}
\label{tab:meas-HI}
\begin{tabular}{lcccrrr}
\hline
Galaxy & RA & Dec & PA & Dist\ \  & $v_{sys}$\ \ \ \  & $v_{max}$\ \ \ \ \\
& & & ($^{\circ}$) & (Mpc) & (km s$^{-1}$) & (km s$^{-1}$)  \\
\hline\hline
ESO074-G015 & 20 52 05.57 & --69 12 05.9 & 141.3\ \  &   6.70 &  590.7 &  93.4 \\
ESO109-G021 & 22 47 06.07 & --64 50 00.6 & 14.9  &  30.40 & 2360.1 & 112.4 \\
ESO115-G021 & 02 37 47.28 & --61 20 12.1 & 43.4  &   4.93 &  512.2 &  64.4 \\
ESO138-G014 & 17 06 59.22 & --62 04 58.3 & 134.4\ \  &  18.57 & 1508.4 & 120.4 \\
ESO146-G014 & 22 13 00.08 & --62 04 05.5 & 47.3  &  21.45 & 1691.1 &  70.2 \\
ESO274-G001 & 15 14 13.84 & --46 48 28.6 & 36.9  &   3.40 &  522.8 &  89.4 \\
ESO435-G025 & 09 59 55.77 & --29 37 01.1 & 74.5  &  29.89 & 2477.0 & 236.3 \\
UGC07321    & 12 17 34.56 & +22 32 26.4 & 81.8  &  10.00 &  401.9 & 112.1 \\
\hline\hline
Galaxy & $W_{20}$ & $W_{50}$ & FI & $M_{\rm HI}$\ \  & \multicolumn{2}{c}{Diam.} \\
 & (km s$^{-1}$) & (km s$^{-1}$) & (Jy km s$^{-1}$) & (M$_{\odot}$)\ \  & (arcmin) & (kpc) \\
\hline\hline
ESO074-G015 & 160.8 & 186.8 &  37.9 & 4.0 10$^{8}$ &   9.350 & 18.2 \\
ESO109-G021 & 211.4 & 224.8 &  10.9 & 2.4 10$^{9}$ &   5.020 & 44.4 \\
ESO115-G021 & 112.2 & 128.9 &  72.0 & 3.4 10$^{8}$ &  13.600 & 19.5 \\
ESO138-G014 & 226.3 & 240.8 &  51.0 & 4.2 10$^{9}$ &   9.210 & 49.8 \\
ESO146-G014 & 127.7 & 140.4 &  \ \ 8.4 & 9.1 10$^{8}$ &   3.840 & 24.0 \\
ESO274-G001 & 167.7 & 178.8 & 152.6\ \  & 4.2 10$^{8}$ &  16.550 & 16.4 \\
ESO435-G025 & 460.9 & 472.7 &  13.1 & 2.8 10$^{9}$ &   7.300 & 63.5 \\
UGC07321    & 209.1 & 224.2 &  38.3 & 9.0 10$^{8}$ &   9.580 & 27.9 \\
\hline
\end{tabular}
\end{table*}

In this first paper we discuss the selection of our sample and present \HI\
observations. In paper II we will review methods to derive the information we
need for our analysis from these data, namely the rotation curves and the \HI\
distributions, \HI\ velocity dispersion and the flaring of the \HI\ layer, all
as a function of radius in the deprojected galaxy plane. Paper III will be
devoted to applying this to our data and presenting the results for each
individual galaxy. In paper IV will analyse the data of one galaxy in our
sample, namely UCG7321 and we will set limits on the flattening of its 
dark halo. 

\section{Observations}
\label{sec:obs-HI}

\subsection{Sample selection}

We selected a sample of southern disk galaxies, for observation
with the Australia Telescope Compact Array (ATCA), that were small,
\HI-rich and late-type, in order to maximise the expected flaring
and the likelihood that the \HI\ brightness would be sufficient to
probe the low surface densities needed to measure the vertical
structure of disk galaxies. 

The galaxies were chosen to be edge-on (optical major-to-minor axis
ratio $a/b \geq 10$) to simplify \HI\ modelling and increase the
accuracy of the measured \HI\ density and kinematics.  We avoided 
galaxies at
Galactic latitude $|b|<10^{\circ}$ to reduce
Galactic extinction in the optical bands. To minimise uncertainties in the
stellar mass distribution we chose relatively bulgeless galaxies
with Hubble type Scd-Sd. The galaxies were
chosen to be nearer than $30$ Mpc to enable the flaring of the \HI\ to be
resolved with the $6$\arcsec\ ATCA synthesised beam. At $30$ Mpc the 
FWHM of the
ATCA beam is $\sim$1 kpc.  The vicinity of each galaxy was searched
in NED for nearby neighbours to ensure that it was isolated. The
\HI\ masses of all galaxies satisfying these criteria were measured
from the \HI\ Parkes All-Sky survey (HIPASS) online data
release\footnote{http://www.atnf.csiro.au/research/multibeam/release}.
With a minimum \HI\ flux integral of $15$ Jy \kms, this limited our
sample to only 5 galaxies. 

The nearest isolated thin edge-on galaxy, ESO274-G001, was added to
the sample. Although it had a Galactic latitude of $9.3^\circ$, it is
exceptionally close at $3.4$ Mpc in the Centaurus A galaxy group,
allowing the \HI\ disk to be measured at a resolution of $\sim$150
pc. A search of \HI\ observations
in the VLA archive revealed the superthin northern Sd galaxy UGC7321
which we also added to our sample. We also added the Sb
galaxy IC2531 to our sample, since it has an extended \HI\ disk and 
a large quantity of archival observations available. The range of
maximum rotation velocity of our sample galaxies in shown in 
column 7 of Table~\ref{tab:meas-HI}.
With the inclusion of IC2531, a galaxy with a box/peanut
bulge, and hence barred \citep{bf1999}, we
attempt to measure the halo shape of galaxies spanning
different mass scales and stages of secular evolution.

\begin{subfigures}
\begin{figure}[t]
\centering
\includegraphics[width=9cm,bbllx=39,bblly=190,bburx=459,
        bbury=587,clip=]{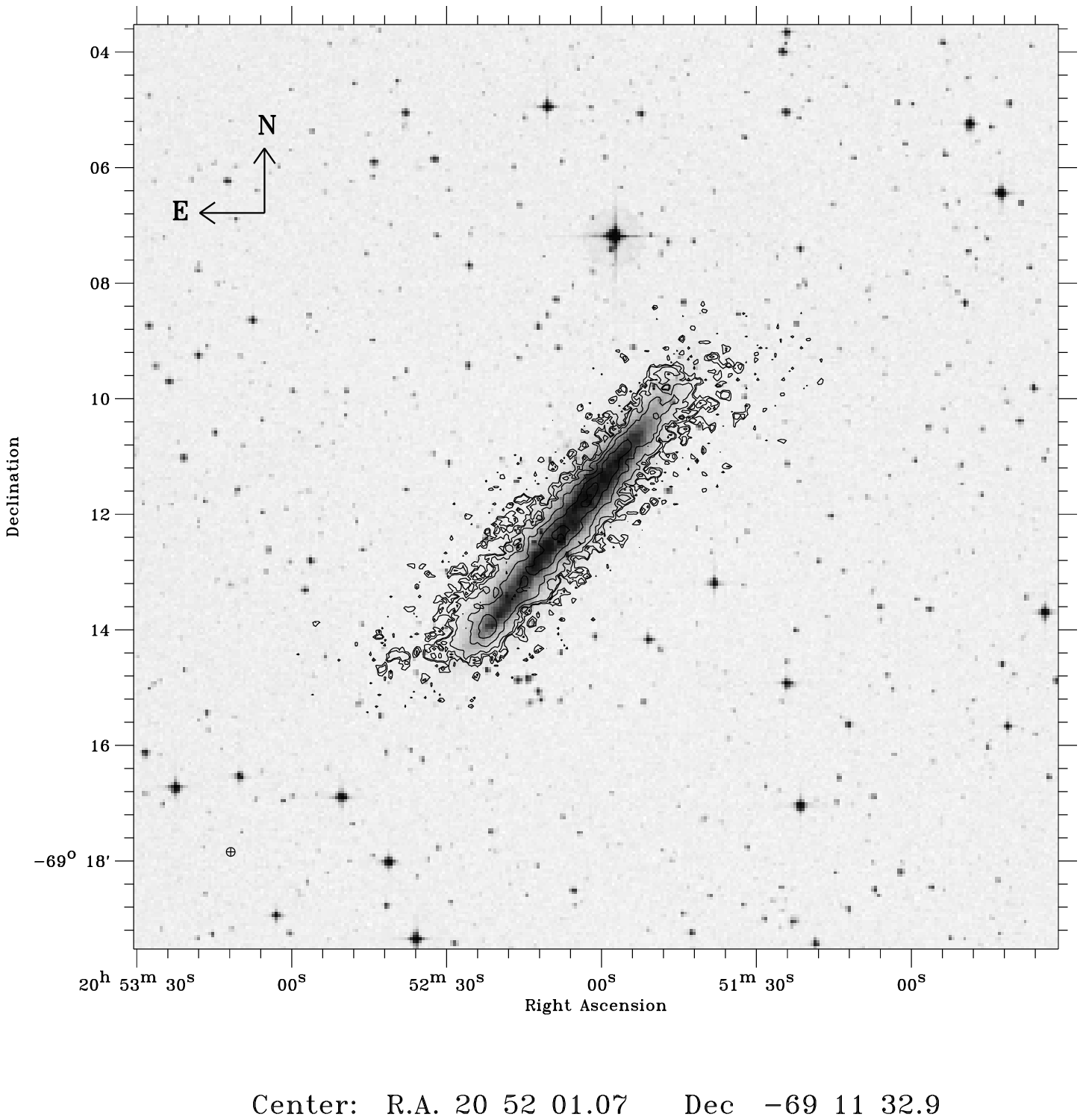}  
\caption{ESO074-G015 (IC5052). \HI\ column density map overlaid on
    the DSS image. Contours are plotted in
    $(3,5,10,25,50,80) \times \sigma$, where the rms noise $\sigma =
    2.05$ \mjybeam or $2.79\times 10^{19}$ \atomscmsq. The FWHM
    synthesised beam has dimensions $9.0\arcsec\times9.0\arcsec$, and is
    displayed as a cross-hatched symbol in the lower left.}
  \label{fig:ch2-eso074-g015-m0}
\end{figure}

\begin{figure}[t]
\centering
\includegraphics[width=9cm]{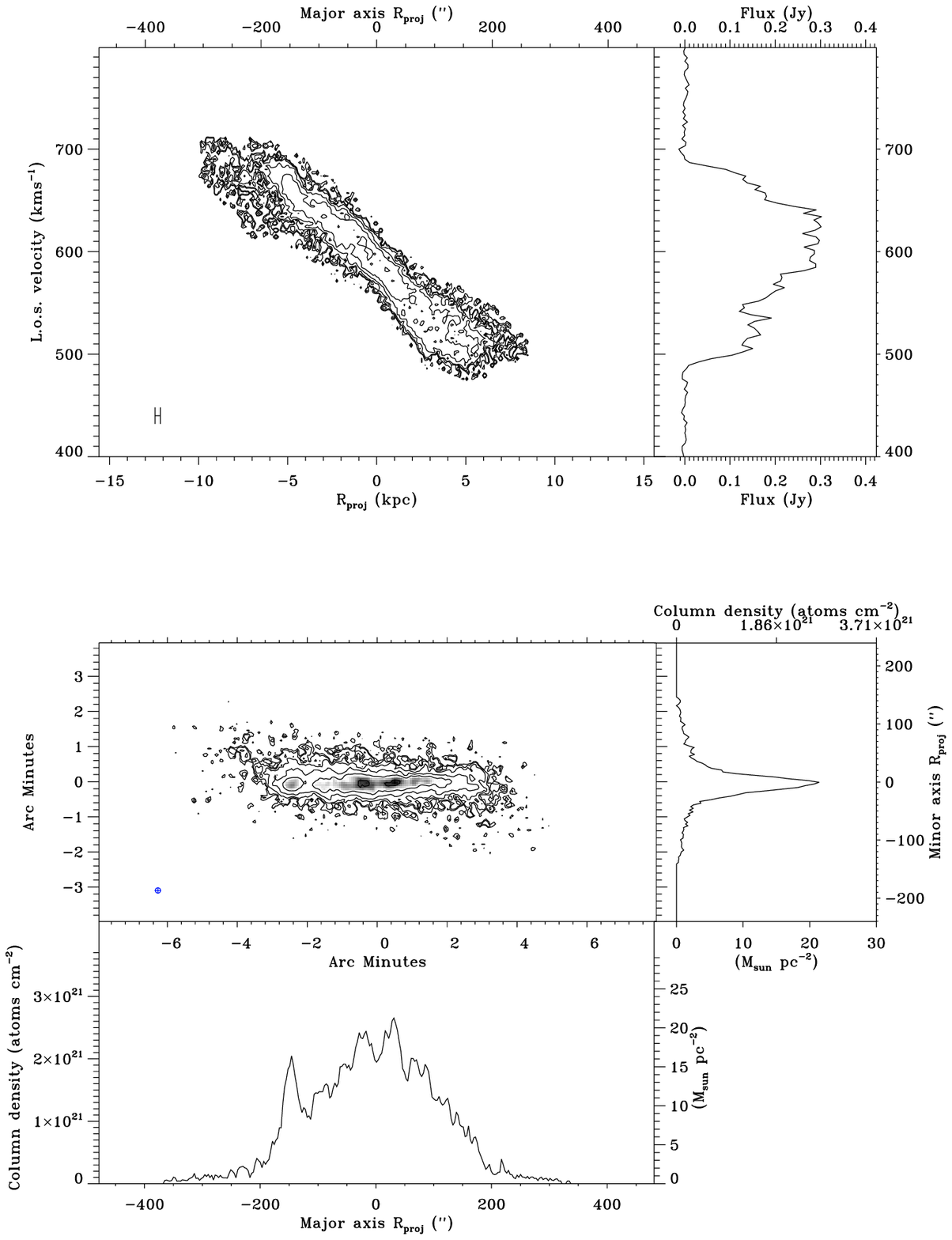} 
  \caption{ESO074-G015 (IC5052). Top left: XV map. XV contours are
    $(3,5,10,25,50,100)\times \sigma$, where the rms noise $\sigma =
    1.90$ \mjybeam or $2.58\times 10^{19}$ \atomscmsq. The half power
    beam extent over the major axis is shown in the lower left corner. 
    Top right: Integrated spectrum. Middle left: \HI\
    column density map rotated with the galaxy major axis aligned with
    the X axis. Column density map contours are $(3,5,10,25,50,100)
    \times \sigma$, where the rms noise $\sigma = 1.65$ \mjybeam\ or
    $2.24\times 10^{19}$ \atomscmsq. The synthesised beam has
    dimensions $9.0\arcsec\times9.0\arcsec$, and is displayed as a
    cross-hatched symbol in the lower left.  Middle right:
    Minor axis profile. Bottom left: Major axis profile.}
  \label{fig:ch2-eso074-g015-combi-plot}
\end{figure}

\addtocounter{figure}{1}
\begin{figure}[t]
\sidecaption
\includegraphics[width=9cm]{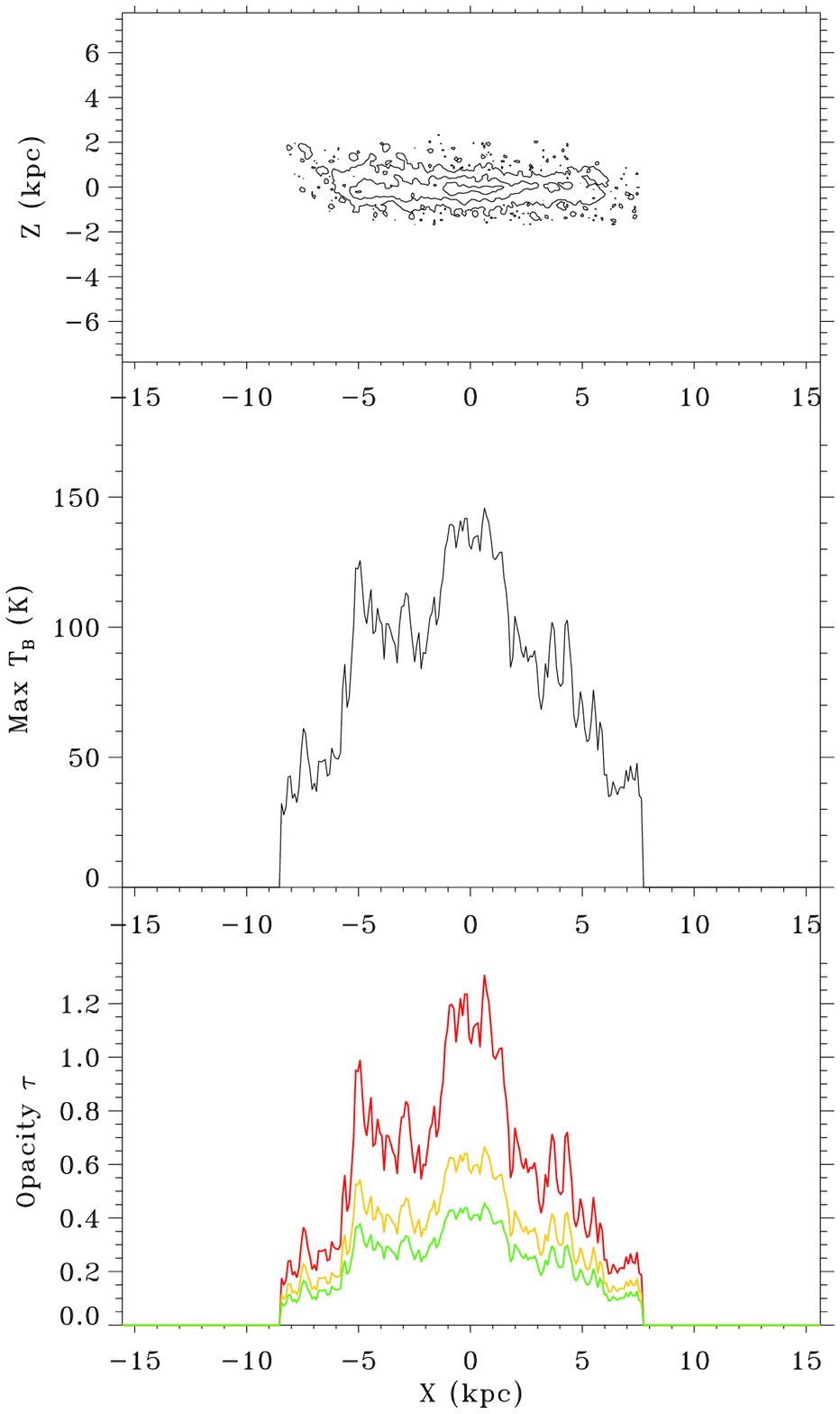}
  \caption{ESO074-G015
    (IC5052). Top : Peak brightness temperature map. Contours are
    plotted in $(35,75,120) \times \sigma$, where the rms noise is
    $10.0$ K. The FWHM synthesised beam has dimensions
    $9.0\arcsec\times9.0\arcsec$.  Middle: Major axis
    peak brightness temperature profile.  Bottom: Inferred \HI\ opacity
    calculated assuming constant \HI\ spin temperatures. 
    The resulting maximum opacities along each line of
    sight column through the galaxy disk are plotted 
    for $T_{spin}$ = 200, 300 and  400 K (bottom to top).}
  \label{fig:ch2-eso074-g015-peaktemp-plot}
\end{figure}
\end{subfigures}

\begin{subfigures}
\begin{figure}[t]
\centering
\includegraphics[width=9cm,bbllx=54,bblly=214,bburx=444,
        bbury=582,clip=]{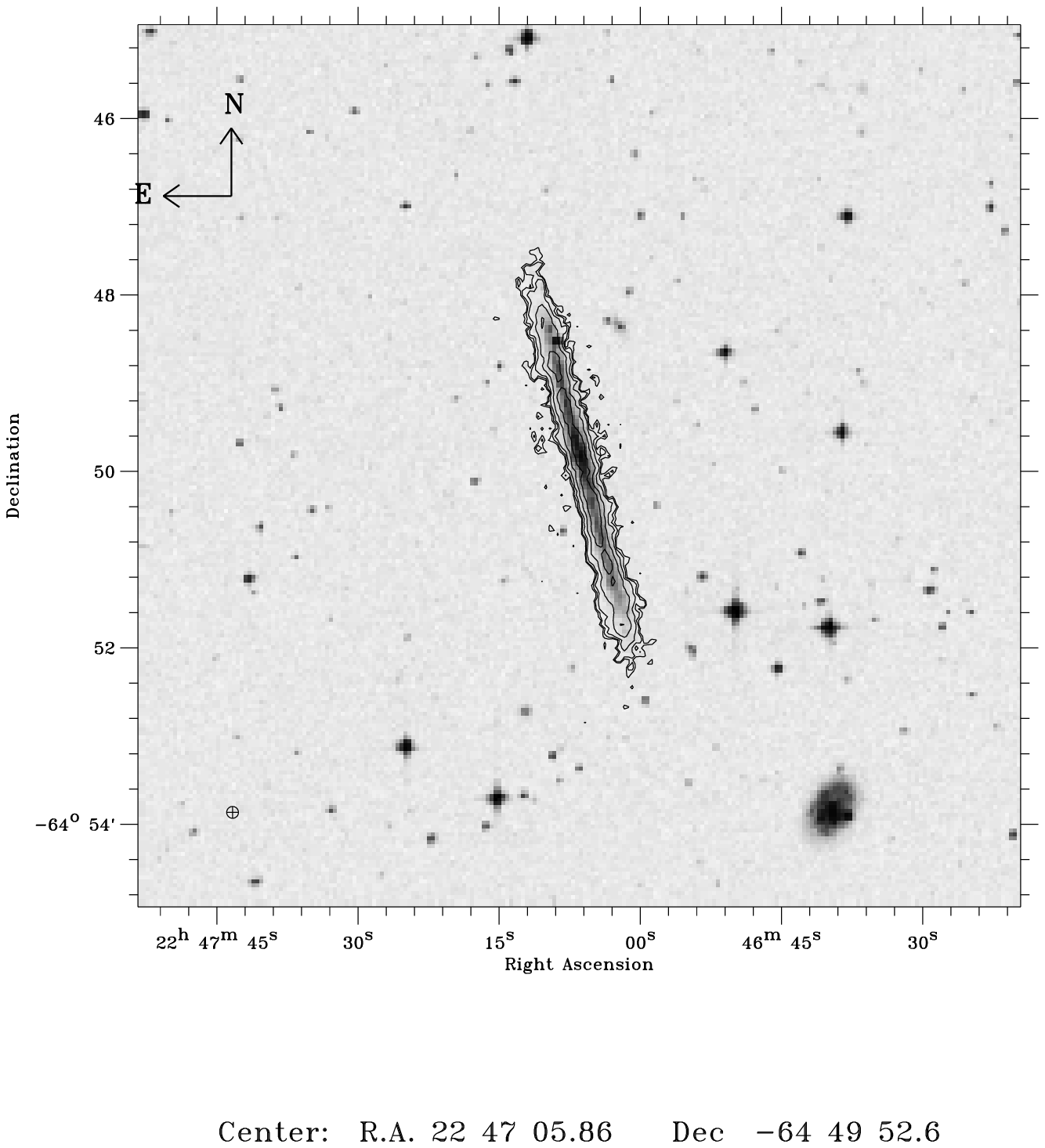}
  \caption[ESO109-G021 (IC5249): \HI\ column density map overlaid on
    the DSS image.]{ESO109-G021 (IC5249). \HI\ column density map overlaid on
    the DSS image. Contours are plotted in
    $(3,5,10,25,50) \times \sigma$, where the rms noise $\sigma =
    1.81$ \mjybeam or $3.11\times 10^{19}$ \atomscmsq. The FWHM
    synthesised beam has dimensions $8.0\arcsec\times8.0\arcsec$, and is
    displayed as a cross-hatched symbol in the lower left.}
  \label{fig:ch2-eso109-g021-m0}
\end{figure}

\addtocounter{figure}{-1}
\begin{figure}[t]
\centering
\includegraphics[width=9cm]{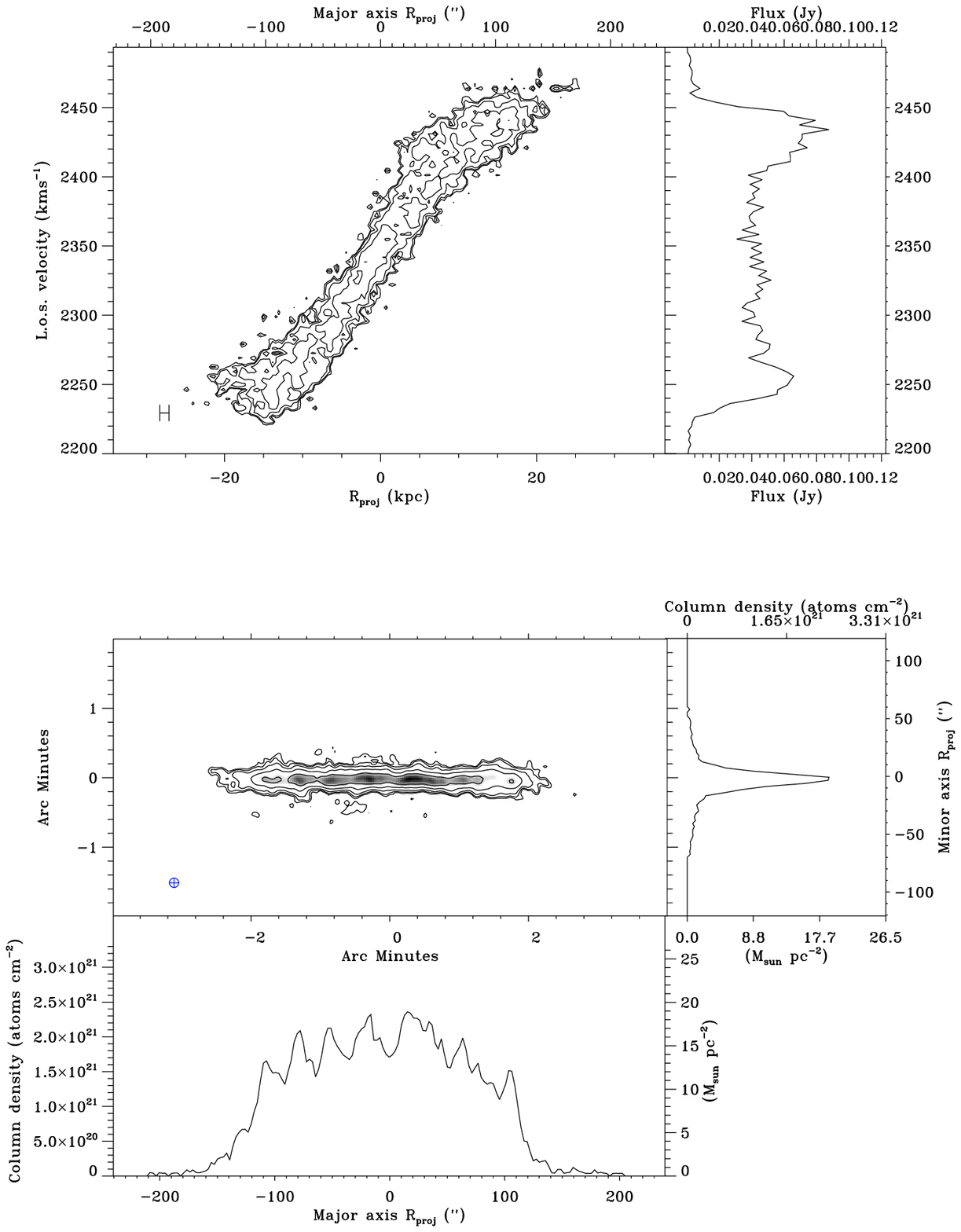}
  \caption[ESO109-G021 (IC5249): XV map, rotated \HI\ column density
  map, integrated \HI\ spectrum and major and minor axis \HI\
  profiles]{ESO109-G021 (IC5249). Top left: XV map. XV contours are
    $(3,5,10,20,40)\times \sigma$, where the rms noise $\sigma = 1.76$
    \mjybeam or $3.03\times 10^{19}$ \atomscmsq. The half power beam
    extent over the major axis is shown in the lower left corner. 
    Top right: Integrated spectrum. Middle left: \HI\ column
    density map rotated with the galaxy major axis aligned with the X
    axis. Column density map contours are $(3,5,10,25,50) \times
    \sigma$, where the rms noise $\sigma = 1.62$ \mjybeam or
    $2.79\times 10^{19}$ \atomscmsq. The FWHM synthesised beam has
    dimensions $8.0\arcsec\times8.0\arcsec$, and is displayed as a
    cross-hatched symbol in the lower left.  Middle right:
    Minor axis profile. Bottom left: Major axis profile.}
  \label{fig:ch2-eso109-g021-combi-plot}
\end{figure}

\begin{figure}[t]
\sidecaption
\includegraphics[width=9cm]{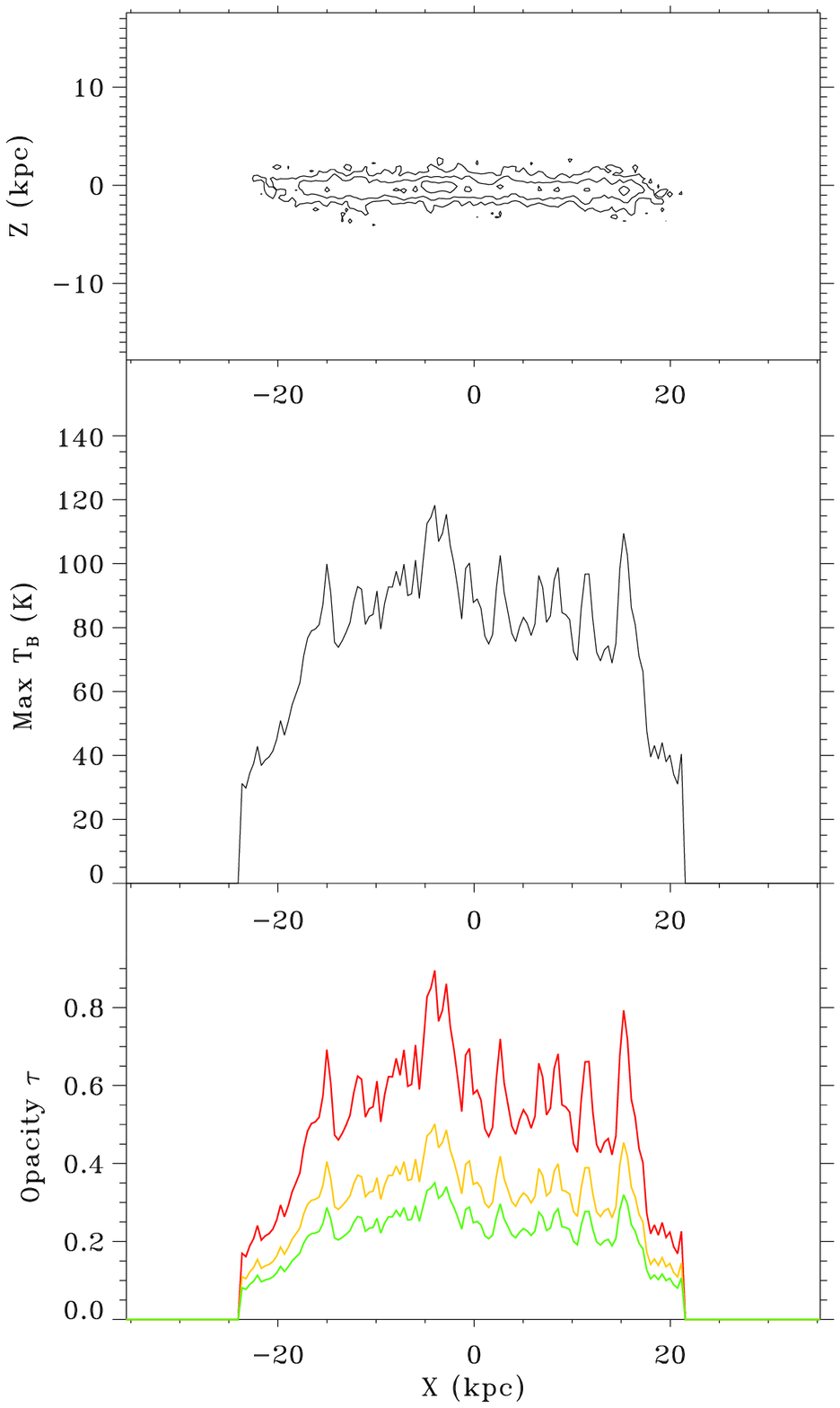}
  \caption[ESO109-G021 (IC5249): \HI\ peak brightness temperature map,
  maximum brightness temperature profile along the major axis, and
  maximum inferred \HI\ opacity along the major axis]{ESO109-G021 (IC5249). 
    Top : Peak brightness temperature
    map. Contours are plotted in $(35,58,93) \times \sigma$,
    where the rms noise is $11.7$ K. The FWHM synthesised beam has
    dimensions $8.0\arcsec\times8.0\arcsec$.  Middle: Major axis
    peak brightness temperature profile.  Bottom: Inferred \HI\ opacity
    calculated assuming constant \HI\ spin temperatures.
    The resulting maximum opacities along each line of
    sight column through the galaxy disk are plotted 
    for $T_{spin}$ = 200, 300 and  400 K (bottom to top).}
  \label{fig:ch2-eso109-g021-peaktemp-plot}
\end{figure}
\end{subfigures}

\subsection{Observing modes and method}

Galaxies were observed with the ATCA in a range of array configurations
to obtain high spatial resolution across each galaxy.  In addition, available
observations from the ATCA and VLA archives
were also used. The list of observations is shown in
Table~\ref{tab:obs-HI}. Column 8 shows the integration time in hours
of each observation. The resulting resolution and
sensitivity of the \HI\ observations are shown in Table~\ref{tab:resol-HI}
and Table~\ref{tab:noise-HI}, respectively. We have smoothed the data in 
Right Ascension to match that in Declination, providing the smallest circular
beam possible. The spatial resolutions shown in Table~\ref{tab:resol-HI} 
correspond to the FWHM of the resulting beams.

At the ATCA the XX and YY polarisations were used, with a spectral
channel width of $3.3$ \kms.  At the VLA the RR and LL polarizations
were used, and the channel width was $5.2$ \kms. Due to the narrow
bandwidth available with the VLA correlator at the time of observations,
multiple overlapping bandpasses were observed to fully span the
velocity width of the target galaxies. For those observations, each
bandpass was calibrated and imaged separately, and the resulting
subcubes glued together along the spectral axis.

\section{Data reduction and imaging}
\label{sec:red-HI}

A detailed account and discussion of the data reduction 
and imaging procedures
is given in the online appendix. Here we restrict ourselves to a brief
synopsis.

The reduction of the \HI\ data was performed with the 
radio interferometry data reduction package {\sc miriad}. The
data were calibrated in the usual manner, using primary and
secondary flux calibrators. Solar and terrestrial interference
peaks in the calibrator scans were inspected, and flagged if
necessary. The bandpass calibration was done using either the 
primary or, in case of sufficient signal-to-noise ratio, the 
secondary calibrator. After calibration, the target data were
inspected for interference, and flagged appropriately.
The radio continuum contribution to each channel map was removed
in the uv-plane, by using a suitable average of line-free channel
maps.

\begin{subfigures}
\begin{figure}[t]
\centering
\includegraphics[width=9cm,bbllx=30,bblly=175,bburx=470,
        bbury=600,clip=]{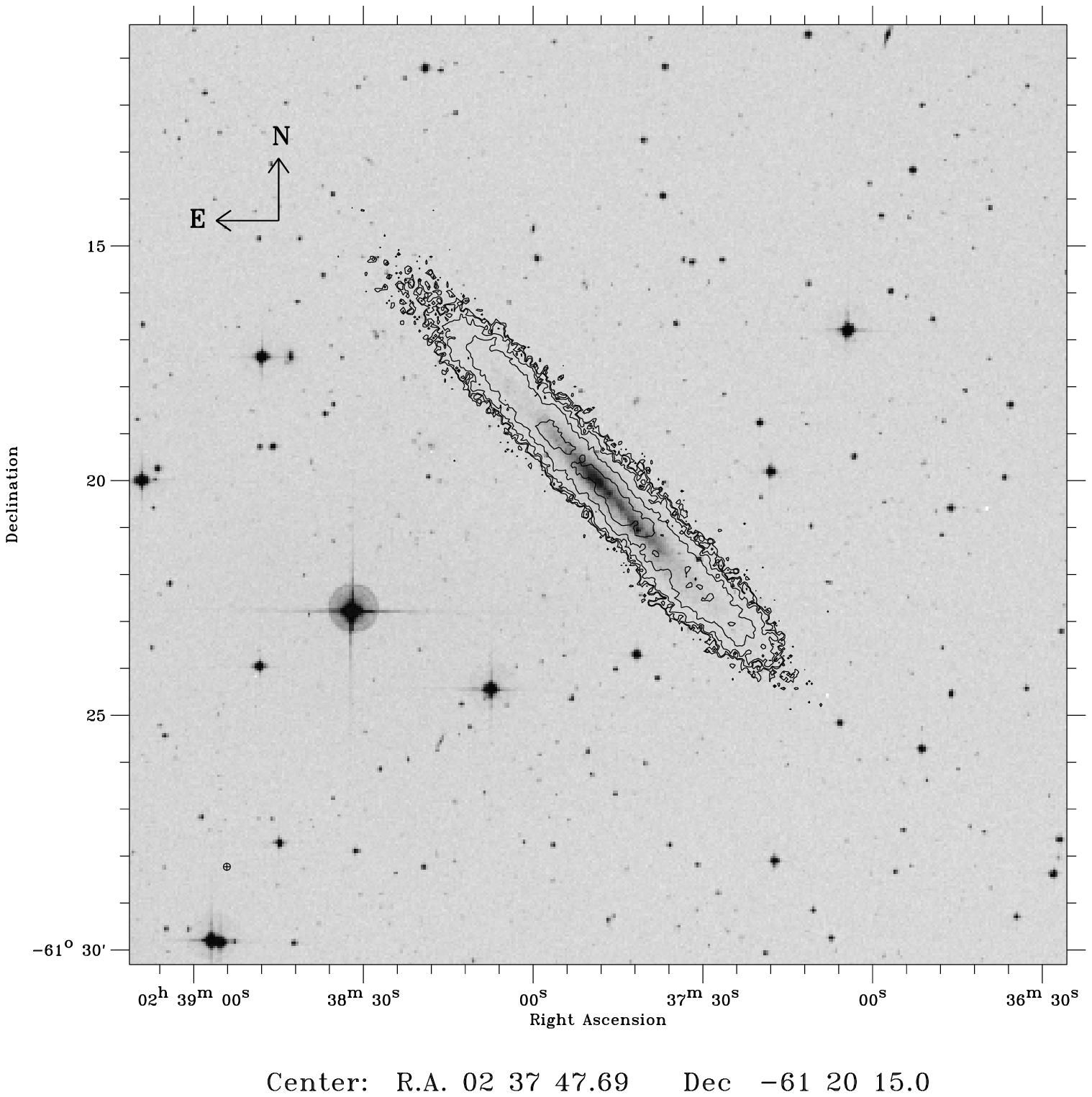}
  \caption[ESO115-G021: \HI\ column density map overlaid on
    the DSS image.]{ESO115-G021. \HI\ column density map overlaid on
    the DSS image. Contours are plotted in
    $(3,5,10,25,50) \times \sigma$, where the rms noise $\sigma =
    1.76$ \mjybeam or $2.44\times 10^{19}$ \atomscmsq. The FWHM
    synthesised beam has dimensions $8.9\arcsec\times8.9\arcsec$, and is
    displayed as a cross-hatched symbol in the lower left.}
  \label{fig:ch2-eso115-g021-m0}
\end{figure}

\addtocounter{figure}{-2}
\begin{figure}[t]
\centering
\includegraphics[width=9cm]{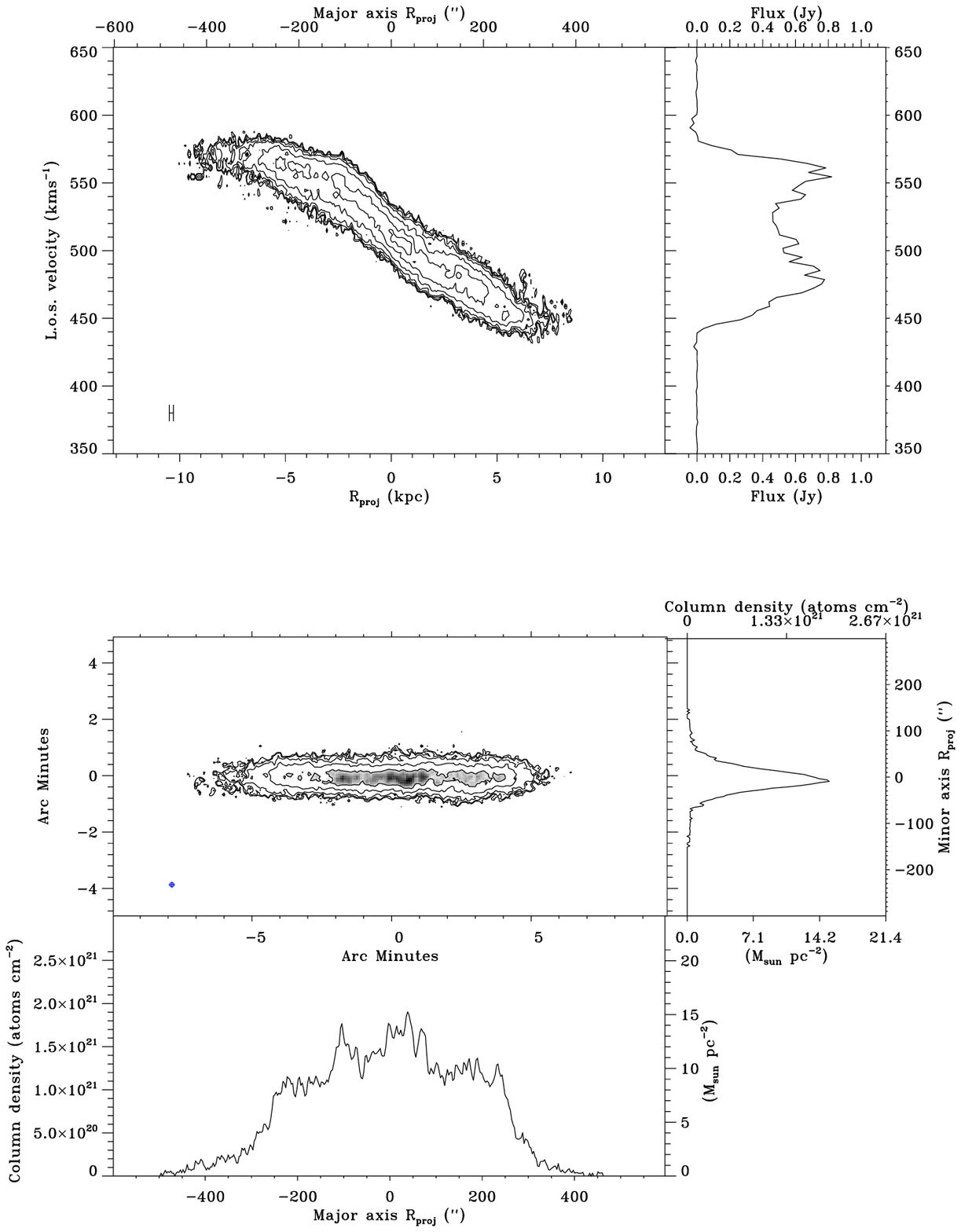}
  \caption[ESO115-G021: XV map, rotated \HI\ column density
  map, integrated \HI\ spectrum and major and minor axis \HI\
  profiles]{ESO115-G021. Top left: XV map. XV contours are
    $(3,5,10,20,50,100,150)\times \sigma$, where the rms noise $\sigma
    = 1.55$ \mjybeam or $2.16\times 10^{19}$ \atomscmsq. The half power
    beam extent over the major axis is shown in the lower left corner.
    Top right: Integrated spectrum. Middle left: \HI\
    column density map rotated with the galaxy major axis aligned with
    the X axis. Column density map contours are $(3,5,10,25,50) \times
    \sigma$, where the rms noise $\sigma = 1.47$ \mjybeam\ or
    $2.04\times 10^{19}$ \atomscmsq. The FWHM synthesised beam has
    dimensions $8.9\arcsec\times8.9\arcsec$, and is displayed as a
    cross-hatched symbol in the lower left.  Middle right:
    Minor axis profile. Bottom left: Major axis profile.}
  \label{fig:ch2-eso115-g021-combi-plot}
\end{figure}

\addtocounter{figure}{-1}
\begin{figure}[t]
\sidecaption
\includegraphics[width=9cm]{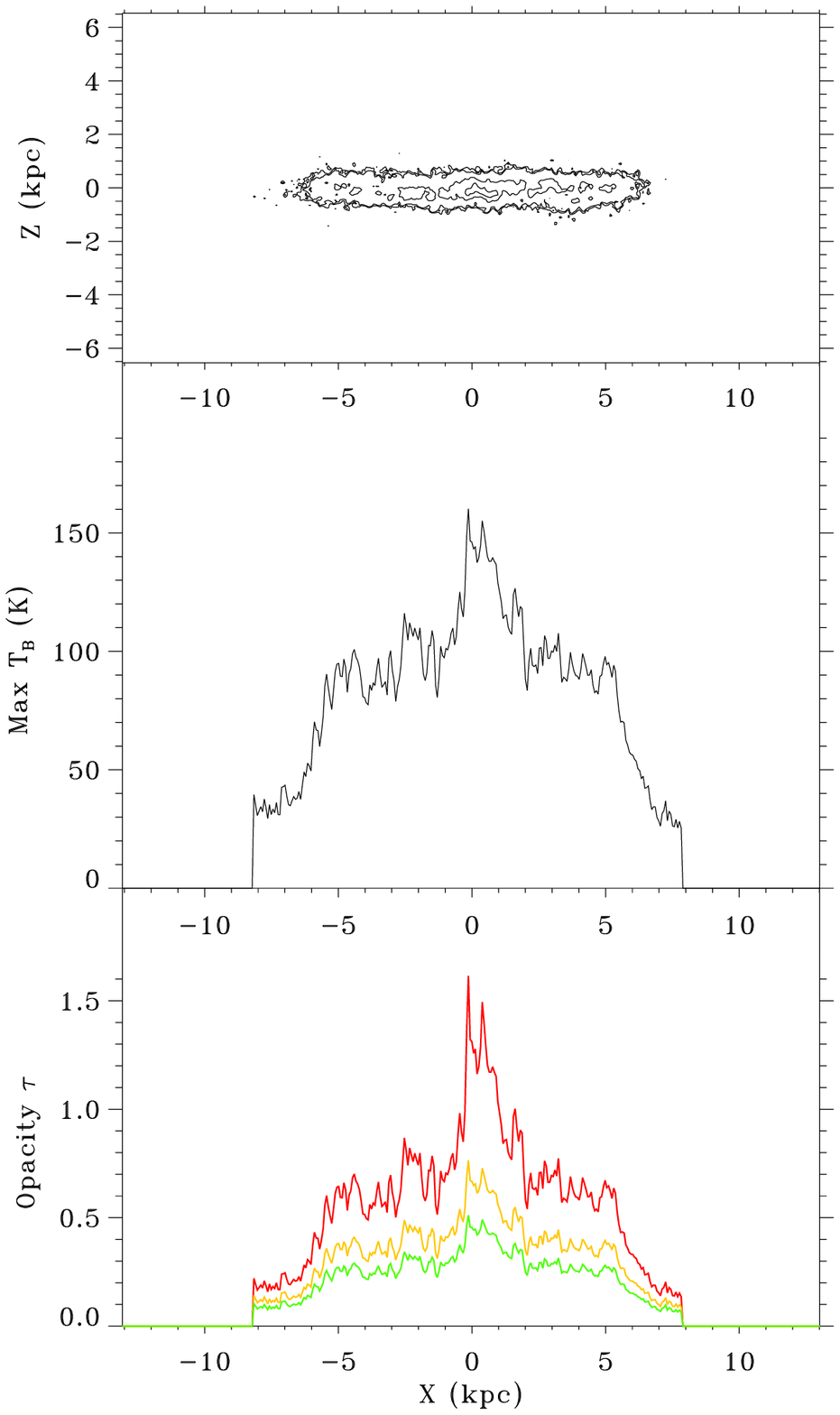}
  \caption[ESO115-G021: \HI\ peak brightness temperature map,
  maximum brightness temperature profile along the major axis, and
  maximum inferred \HI\ opacity along the major axis]{ESO115-G021. 
Top : Peak brightness temperature map. Contours 
    are plotted in $(35,42,86,129) \times \sigma$, where the rms noise
    is $8.6$ K. The FWHM synthesised beam has dimensions
    $8.9\arcsec\times8.9\arcsec$.  Middle: Major axis
    peak brightness temperature profile.  Bottom: Inferred \HI\ opacity
    calculated assuming constant \HI\ spin temperatures. 
    The resulting maximum opacities along each line of
    sight column through the galaxy disk are plotted
    for $T_{spin}$ = 200, 300 and  400 K (bottom to top).}
  \label{fig:ch2-eso115-g021-peaktemp-plot}
\end{figure}
\end{subfigures}

\begin{subfigures}
\begin{figure}[t]
\centering
\includegraphics[width=9cm,bbllx=80,bblly=255,bburx=415,
        bbury=565,clip=]{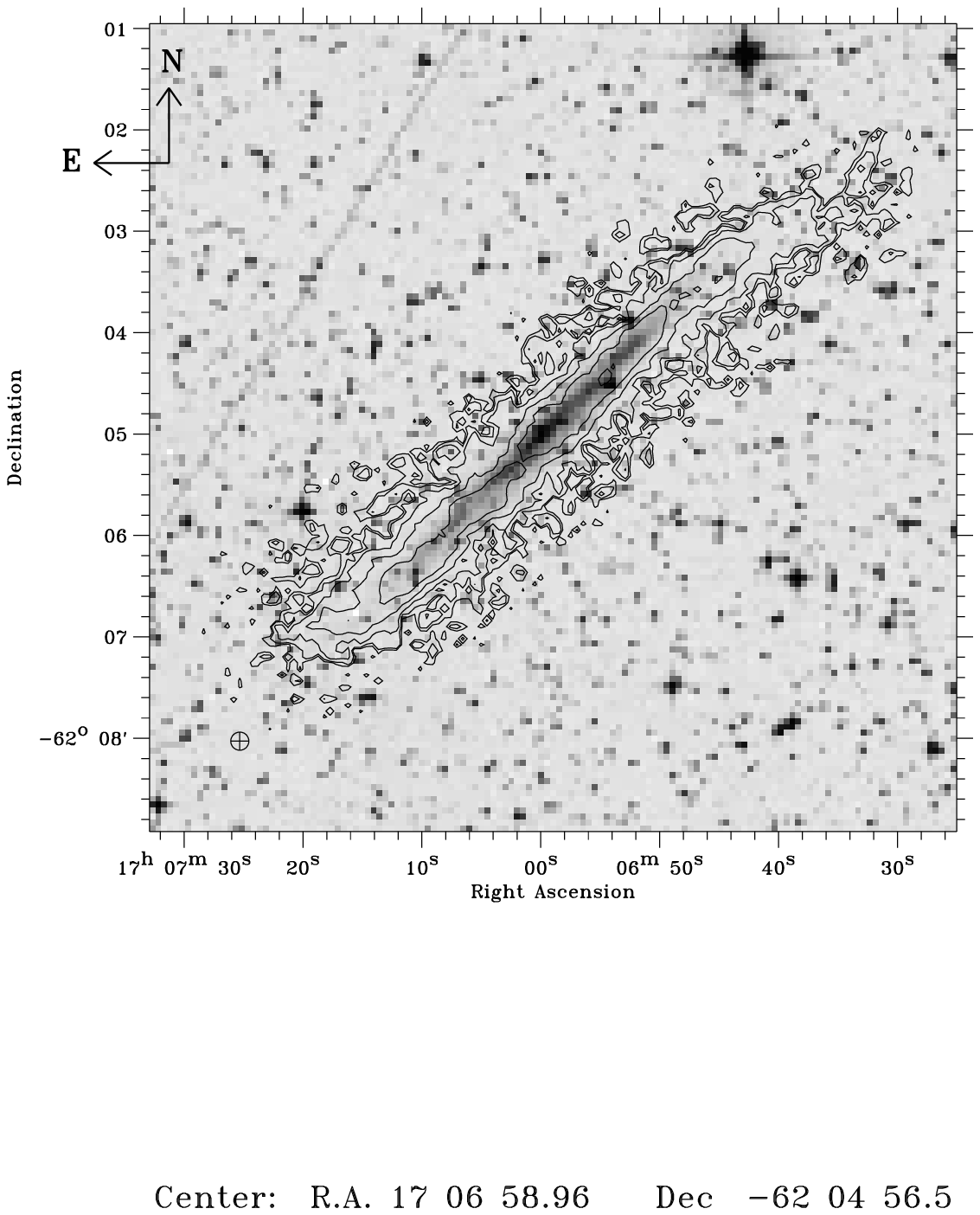}
  \caption[ESO138-G014: \HI\ column density map overlaid on
    the DSS image.]{ESO138-G014. \HI\ column density map overlaid on
    the DSS image. Contours are plotted in
    $(3,5,10,25,50,100) \times \sigma$, where the rms noise $\sigma =
    2.84$ \mjybeam or $2.73\times 10^{19}$ \atomscmsq. The FWHM
    synthesised beam has dimensions $10.7\arcsec\times 10.7\arcsec$, and is
    displayed as a cross-hatched symbol in the lower left.}
  \label{fig:ch2-eso138-g014-m0}
\end{figure}

\addtocounter{figure}{-3}
\begin{figure}[t]
\centering
\includegraphics[width=9cm]{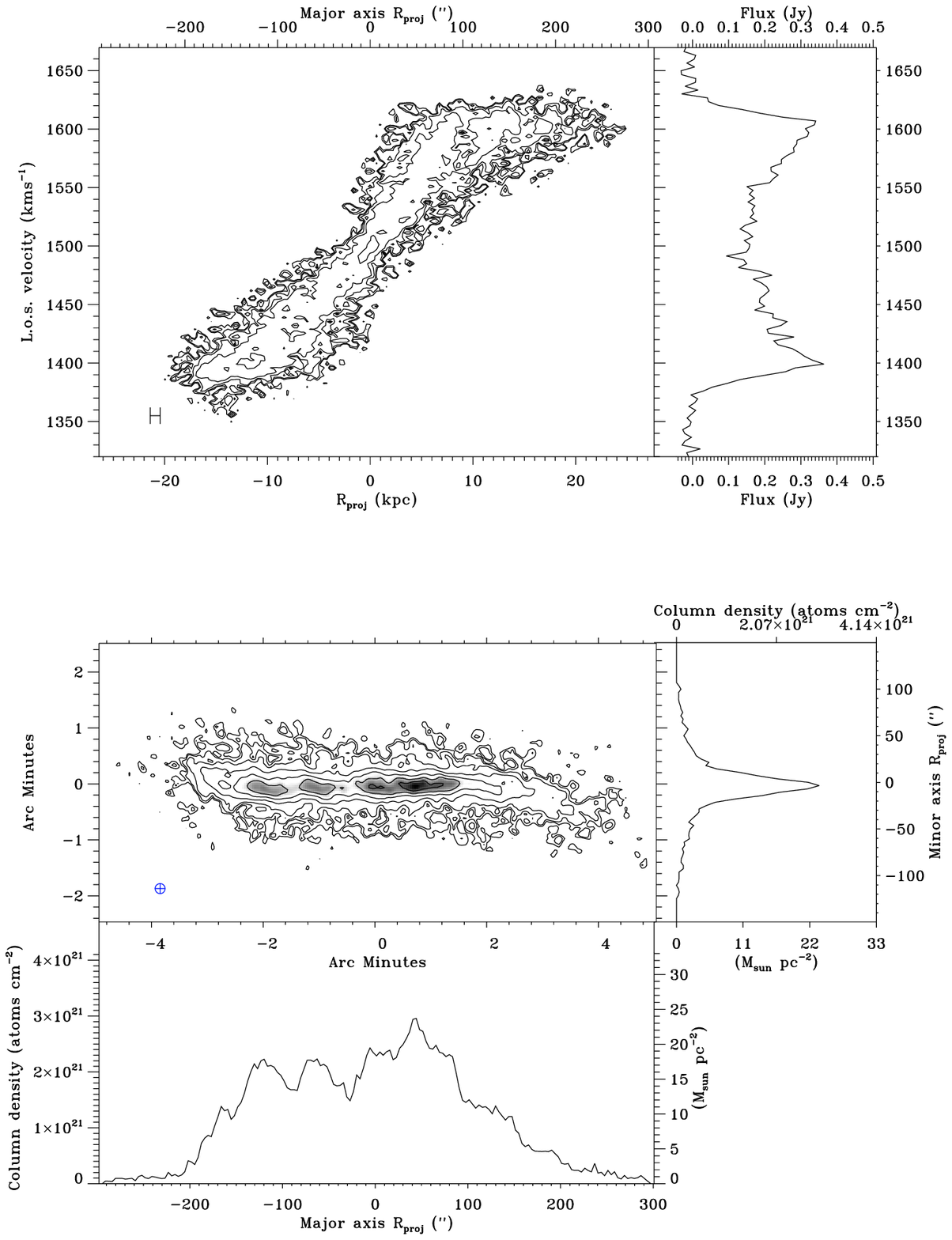}
  \caption[ESO138-G014: XV map, rotated \HI\ column density
  map, integrated \HI\ spectrum and major and minor axis \HI\
  profiles]{ESO138-G014. Top left: XV map. XV contours are
    $(3,5,10,20,50)\times \sigma$, where the rms noise
    $\sigma = 2.74$ \mjybeam\ or $2.64\times 10^{19}$ \atomscmsq. The
    half power beam extent over the major axis is shown in the lower
    left corner. Top right: Integrated spectrum. Middle
    left: \HI\ column density map rotated with the galaxy major axis
    aligned with the X axis. Column density map contours are
    $(3,5,10,25,50,75,100) \times \sigma$, where the rms noise $\sigma
    = 2.38$ \mjybeam\ or $2.29\times 10^{19}$ \atomscmsq. The FWHM
    synthesised beam has dimensions $10.7\arcsec\times 10.7\arcsec$, and is
    displayed as a cross-hatched symbol in the lower left.
    Middle right: Minor axis profile. Bottom left: Major axis
    profile.}
  \label{fig:ch2-eso138-g014-combi-plot}
\end{figure}

\addtocounter{figure}{-2}
\begin{figure}[t]
\sidecaption
\includegraphics[width=9cm]{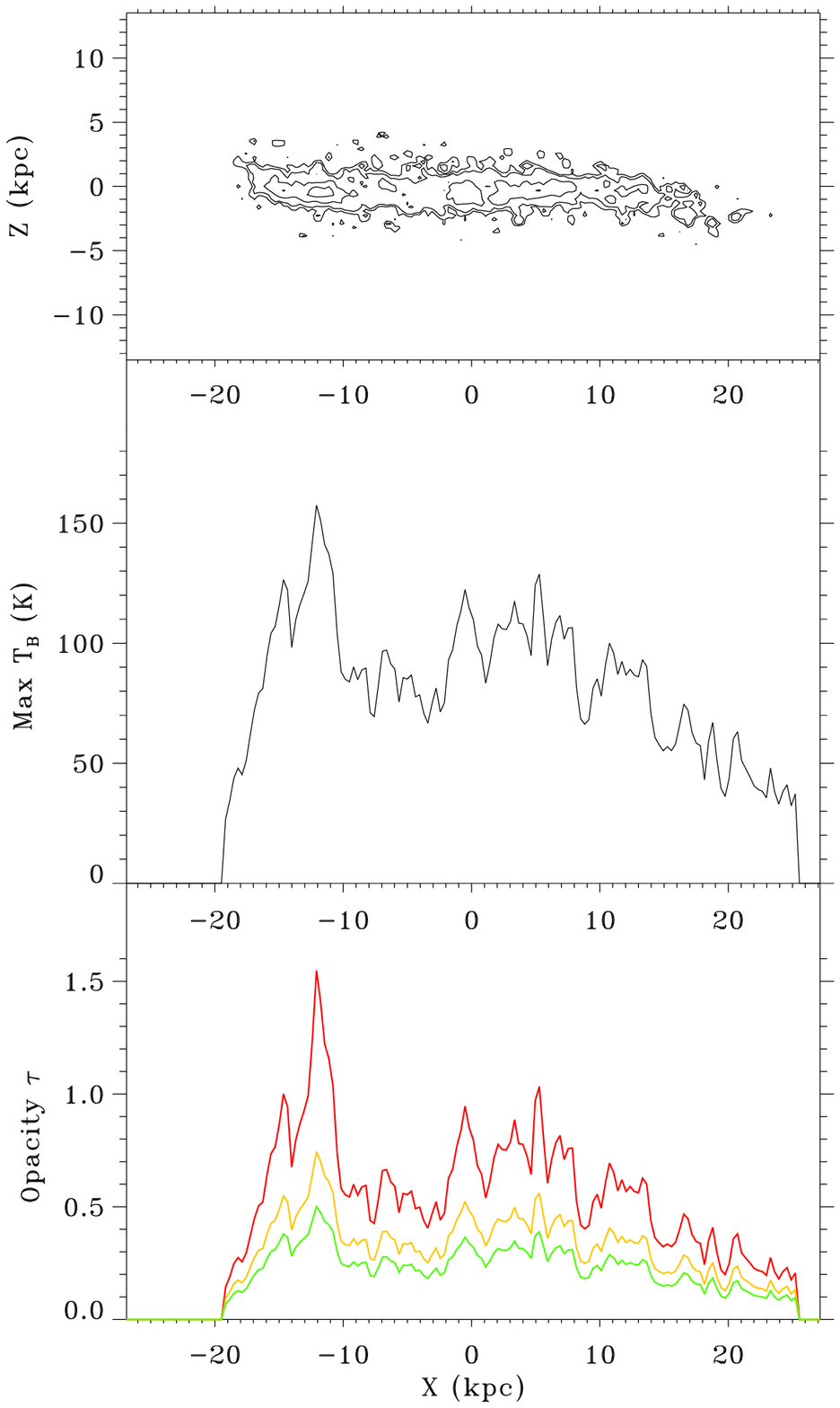}
  \caption[ESO138-G014: \HI\ peak brightness temperature map,
  maximum brightness temperature profile along the major axis, and
  maximum inferred \HI\ opacity along the major axis]{ESO138-G014. 
Top : Peak brightness temperture map. Countours
    are plotted in $(41,51,83,124) \times \sigma$, where the rms noise
    is $10.4$ K. The FWHM synthesised beam has dimensions
    $10.7\arcsec\times 10.7\arcsec$r.  Middle: Major axis
    peak brightness temperature profile.  Bottom: Inferred \HI\ opacity
    calculated assuming constant \HI\ spin temperatures. 
    The resulting maximum opacities along each line of
    sight column through the galaxy disk are plotted 
    for $T_{spin}$ = 200, 300 and  400 K (bottom to top).}
  \label{fig:ch2-eso138-g014-peaktemp-plot}
\end{figure}
\end{subfigures}

\section{\HI\ emission}
\label{sec:emission-HI}

\subsection{The \HI\ distribution}
\label{sec:dist-HI}

To make the \HI\ column density map, the intensity of each channel map
was first converted from flux density $I(x,y)$ in units of Jy beam$^{-1}$ to
\HI\ column density $N_{\rm HI}$ in atoms cm$^{-2}$ using
\begin{equation}
N_{\rm HI}(v) = \left(
  \frac{1.101974351\times 10^{24}}{\theta_{maj}\,\theta_{min}} \right) I(x,y),
\end{equation}
where $\theta_{maj}$ and $\theta_{min}$ are the major and minor axis
FWHM of the synthesised beam in arcsec.  
The \HI\ column density maps were made by taking the zeroth moment of masked 
channel maps (with $v$ in \kms):
\begin{equation}
N_{\rm HI} = \int N_{\rm HI}(v) dv.
\end{equation}

To exclude any residual sidelobes, a loose region mask was defined
interactively for each channel map and then all pixels with flux
density above a nominal clip level of 2-3$\sigma$ in each masked
channel map were integrated.  This procedure ignores faint, low-level 
\HI\ emission and the resulting integrated \HI\ maps will not contain
any extended, low surface brightness component in the gas distribution. However,
this does not have an effect on the general distribution of \HI\ as determined
by our method. The maps in this paper 
have been produced to show the general distribution of the gas 
in the disk in order to judge the suitability of the galaxies for further study. 
We will only be able to analyse the flaring and
velocity dispersion as
long as we can obtain high signal-to-noise profiles. A search 
for extended \HI\ is beyond the scope of our present analysis of the data. 
As the number of channels integrated to
form the column density map varies as a function of position, the
noise in the \HI\ column density map also varies accordingly. In
order to approximate the noise of the \HI\ column density map, a noise
map was also constructed.  The noise at each position in the map was
calculated as
\begin{equation}
\sigma_{total} = \sqrt{n_{\rm ch}} \times \langle \sigma_{\rm ch}(v) \rangle,
\end{equation}
where $\langle \sigma_{\rm ch}(v) \rangle$ is the mean rms noise of all
channel maps integrated at that position. By dividing the column
density map by the noise map a signal-to-noise ratio map was
formed. Following the method of \citet{kvdkdb2004}, the rms noise of
the column density map was calculated from the average flux density of
all pixels with a signal-to-noise ratio between 2.75 and 3.25.

\subsection{Global properties}
\label{sec:global-HI}

The integrated \HI\ spectrum was measured from the \HI\ data cube using
the \HI\ column density map to mask the cube, and integrating over the
area of the galaxy mask. This flux density spectrum was converted to
units of Jy by dividing by the synthesised beam area
\begin{equation}
S(v) = \frac{1}{1.133\,\theta_{maj}\,\theta_{min}}\int I(x,y) dx dy.
\end{equation}
As the flux was integrated over an equal area in each channel map, the
noise of each channel is
\begin{equation}
\sigma_{tot} = \sqrt{{\rm Area}_{\rm HI}} \times \sigma_{\rm ch},
\end{equation}
where Area$_{\rm HI}$ is the area of \HI\ column density emission, and
$\sigma_{\rm ch}$ is the rms noise of each channel map. For \HI\ cubes with
a constant rms noise in each channel, the noise of the spectrum was
fitted from the line-free channels in the spectrum.  The velocity
widths at the $20\%$- and $50\%$-levels ($W_{20}$, $W_{50}$,
respectively) were measured from the spectrum, and the spectrum was
integrated over the velocity range spanned by $W_{20}$ to get the
integrated flux, FI:
\begin{equation}
{\rm FI} = \int^{W_{20,high}}_{W_{20,low}} S(v) dv.
\end{equation}
The restriction of the integration to the velocity range $W_{20}$ introduces
a small but systematic error in the total flux, underestimating
it by no more than a few \%. However, for our present purposes it is important
to examine the integrated profile in order to identify major asymmetries that
would make the galaxy unsuitable for our purposes.

The total \HI\ mass, $M_{\rm HI}$ was then measured from the flux integral
using the formula
\begin{equation}
M_{\rm HI} = 2.35 D^2 \times {\rm FI},
\end{equation}
where the adopted distance $D$ is given in Mpc. These \HI\ measurements
are shown in Table~\ref{tab:meas-HI}. These total masses are also
underestimated by a few \%.

The centre of each galaxy was obtained from the \HI\ column density map by
rotating the image by $180^{\circ}$ and finding the optimal pixel
offset of the rotated image relative to the original image by 
maximising the correlation function of the two images using the IDL
Astronomy User's Library function {\em
  correl\_optimize}.\footnote{http://idlastro.gsfc.nasa.gov/homepage.html.}
Formally this gives the center of the \HI\ disk rather than the
galactic rotational center.  
Comparison to the coordinates given in the NASA/IPAC Extragalactic Database
NED\footnote{nedwww.ipac.caltech.edu.}, which are derived from
the Two Micron All Sky Survey 2MASS\footnote{www.ipac.caltech.edu/2mass/},
show differences of on average only 4 arcsec per coordinate, except for
ESO074-G015, where it is about 20 arcsec in RA and 30 arcsec in Dec. However,
this galaxy shows some deviations from symmetry, both in optical
appearance as well as in the \HI\ distribution. The position adopted 
fits the symmetry of the XV map very well (see 
Fig.~\ref{fig:ch2-eso074-g015-combi-plot}) and is also closer to centers
listed in NED that are derived from photographic and IRAS data.

The position angle of the galaxy was found by a similar method.
First a wide range of position angles were trialed around the
estimated position angle. The \HI\ column density map was reflected
over each test position angle about the galaxy center and the
residual of the two maps was calculated. The position angle that
yielded the lowest summed residual was then adopted as the next
position angle estimate. Subsequent rounds spanned smaller and
smaller ranges of the position angle until the position angle was
determined to less than $0.1\deg$ or until the total residual dropped
below $1\%$ of the summed flux of the \HI\ column density map. Both
methods worked very well for observations with a near circular
synthesised beam, achieving an accuracy of $\sim$1 pixel in the
centre position and $0.1-0.2\deg$ in position angle. However,
observations with a highly elongated beam were poorly fitted, as
the distortion of the image caused by the beam shape biased the
derived PA towards the PA of the synthesised beam.  For such
observations the uncertainty of the measured centre and position
angle were larger.

The {\em xzv} cube was constructed by rotating the \HI\ data cube 
about the galaxy center to align the galaxy major axis with the X axis. In
Sect.~\ref{sec:ch2-results} we show the rotated channel maps to
allow the flaring to be viewed directly from the observations. We also
present the \HI\ column density map with the projected major and minor
axis profiles as measured from the \HI\ column density map. The \HI\
diameter and maximum vertical extent were measured at the $3\sigma$
level of the projected profiles.

An XV cube was also produced by reordering the cube axes to form a
{\em xvz} cube. This cube was then integrated over the $z$ axis to
make an XV map using the same method as was used to make the \HI\
column density map. Both the XV map and the \HI\ column density map
are shown with their respective noise levels in
Sect.~\ref{sec:ch2-results}.  The systemic velocity of the galaxy
was measured from the XV map by shifting the XV distribution flipped
in V with respect to the actual XV map to maximise the correlation
function.

Peak brightness temperature maps were also made for each galaxy,
where the peak brightness temperature is the maximum brightness
temperature over all channel maps at each spatial position. The
brightness temperature $T_B$ was derived from the \HI\ column density
$N_{\rm HI}$ in each channel map using \begin{equation} T_B =
\frac{N_{\rm HI}}{1.83\times 10^{18}}.  \end{equation} The major
axis profile of the peak brightness map is also shown for each
galaxy in Sect.~\ref{sec:ch2-results}. These plots are useful
indicators of possible \HI\ self-absorption that could be obscuring
the intrinsic \HI\ surface distribution. Sect.~\ref{sec:ch2-results}
also shows the inferred \HI\ opacity of the brightest \HI\ emission
at each major axis position, assuming three different values of the
\HI\ spin temperature ($200, 300, 400$ K).

\begin{subfigures}
\begin{figure}[t]
\centering
\includegraphics[width=9cm,bbllx=60,bblly=225,bburx=435,
        bbury=575,clip=]{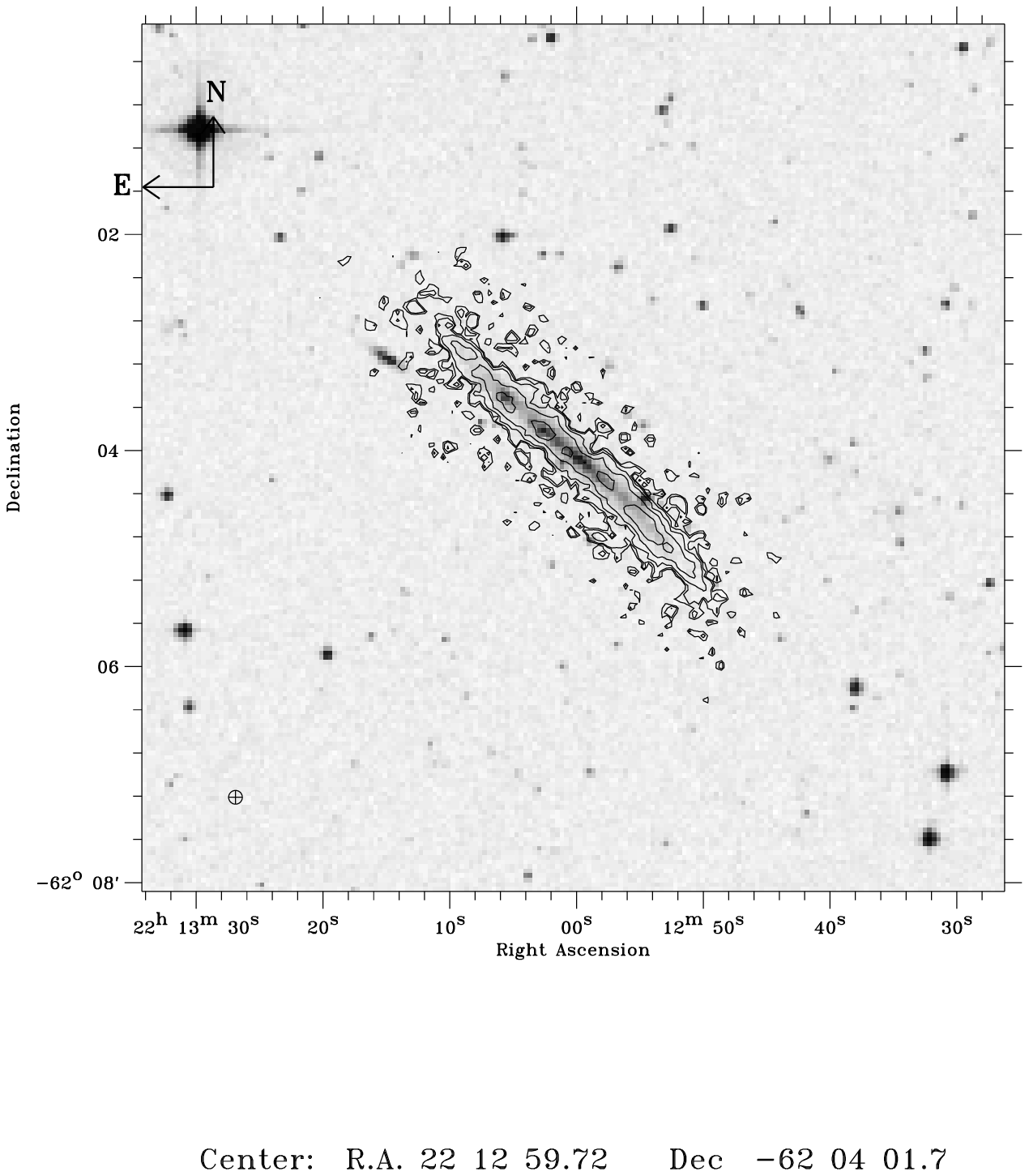}
  \caption[ESO146-G014: \HI\ column density map overlaid on
    the DSS image.]{ESO146-G014. \HI\ column density map overlaid on
    the DSS image. Contours are plotted in
    $(3,5,10,25,50) \times \sigma$, where the rms noise $\sigma =
    1.53$ \mjybeam\ or $2.92\times 10^{19}$ \atomscmsq. The FWHM
    synthesised beam has dimensions $7.6\arcsec\times7.6\arcsec$, and is
    displayed as a cross-hatched symbol in the lower left.}
  \label{fig:ch2-eso146-g014-m0}
\end{figure}

\addtocounter{figure}{-4}
\begin{figure}[t]
\centering
\includegraphics[width=9cm]{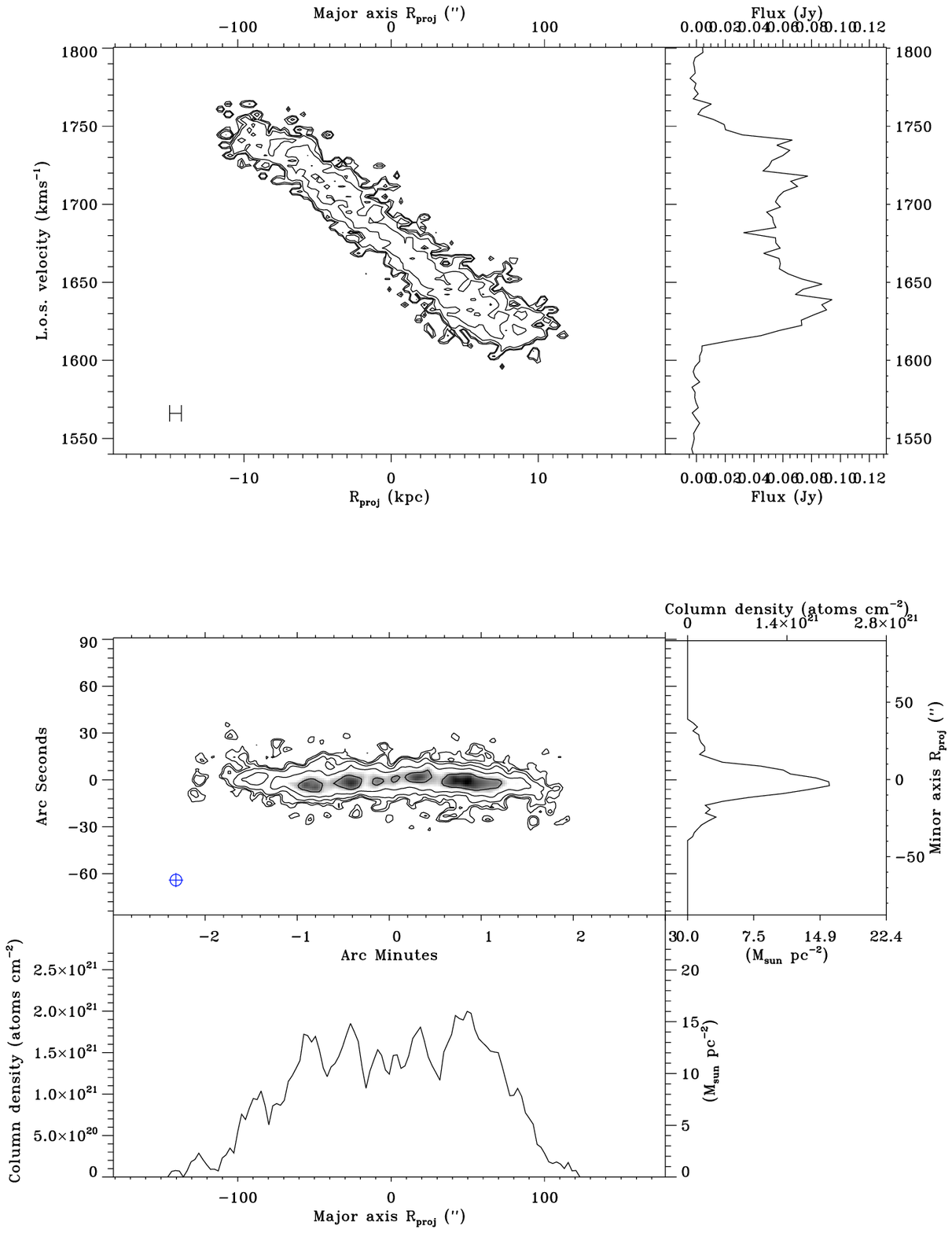}
  \caption[ESO146-G014: XV map, rotated \HI\ column density
  map, integrated \HI\ spectrum and major and minor axis \HI\
  profiles]{ESO146-G014. Top left: XV map. XV contours are
    $(3,5,10,25,50)\times \sigma$, where the rms noise
    $\sigma = 1.45$ \mjybeam\ or $2.76\times 10^{19}$ \atomscmsq. The
    half power beam extent over the major axis is shown in the lower
    left corner. Top right: Integrated spectrum. Middle
    left: \HI\ column density map rotated with the galaxy major axis
    aligned with the X axis. Column density map contours are
    $(3,5,10,25,50) \times \sigma$, where the rms noise $\sigma
   = 1.40$ \mjybeam\ or $2.67\times 10^{19}$ \atomscmsq. The FWHM
    synthesised beam has dimensions $7.6\arcsec\times7.6\arcsec$, and is
    displayed as a cross-hatched symbol in the lower left.
    Middle right: Minor axis profile. Bottom left: Major axis
    profile.}
  \label{fig:ch2-eso146-g014-combi-plot}
\end{figure}

\addtocounter{figure}{-3}
\begin{figure}[t]
\sidecaption
\includegraphics[width=9cm]{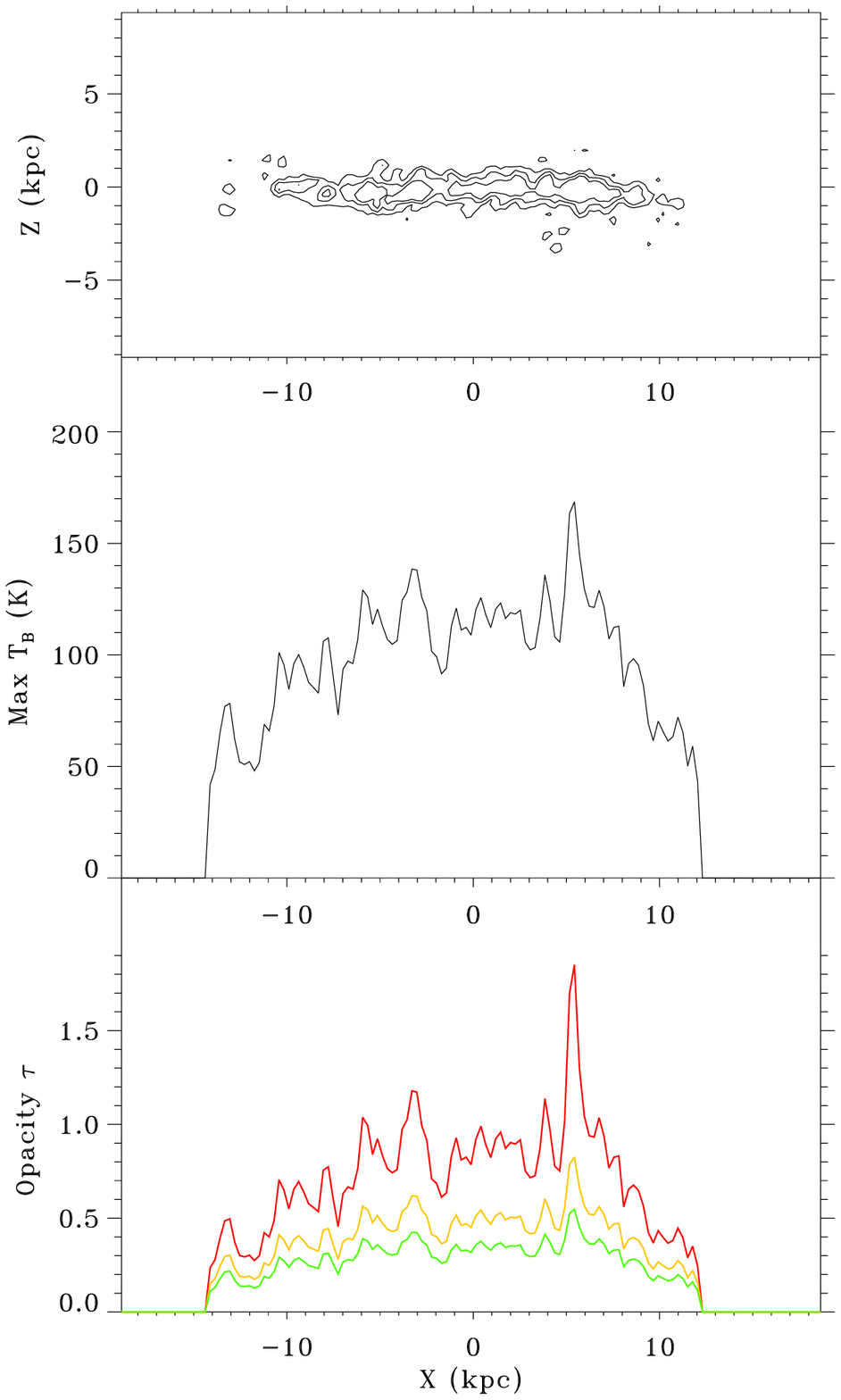}
  \caption[ESO146-G014: \HI\ peak brightness temperature map,
  maximum brightness temperature profile along the major axis, and
  maximum inferred \HI\ opacity along the major axis]{ESO146-G014.
    Top : Peak brightness temperature map. Contours
    are plotted in $(60,80,100) \times \sigma$, where the rms noise
    is $14.9$ K. The FWHM synthesised beam has dimensions
    $7.6\arcsec\times7.6\arcsec$.  Middle: Major axis
    peak brightness temperature profile.  Bottom: Inferred \HI\ opacity
    calculated assuming constant \HI\ spin temperatures. 
    The resulting maximum opacities along each line of
    sight column through the galaxy disk are plotted 
    for $T_{spin}$ = 200, 300 and  400 K (bottom to top).}
  \label{fig:ch2-eso146-g014-peaktemp-plot}
\end{figure}
\end{subfigures}

\begin{subfigures}
\begin{figure}[t]
\centering
\includegraphics[width=9cm,bbllx=20,bblly=167,bburx=480,
        bbury=605,clip=]{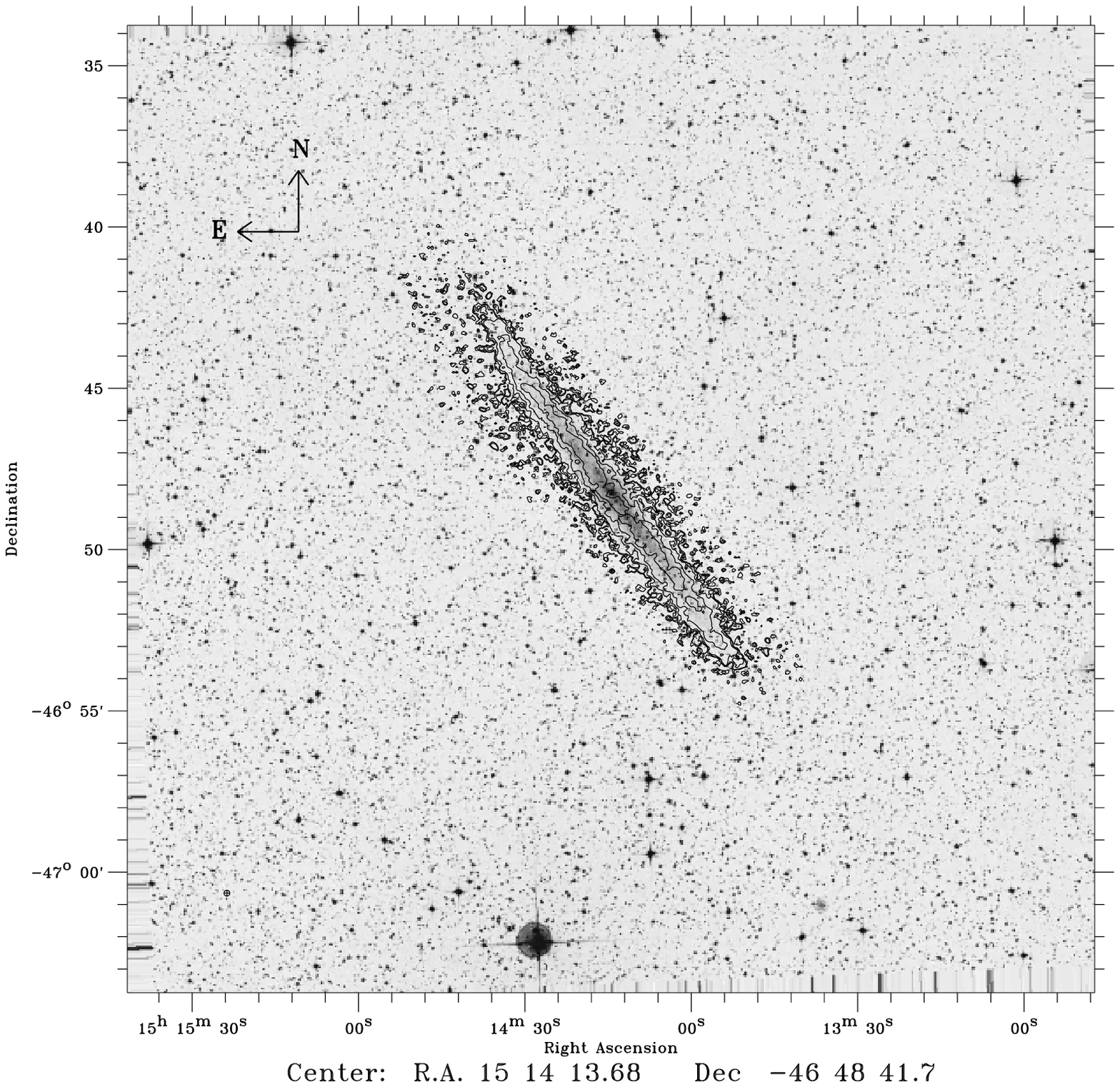}
  \caption[ESO274-G001: \HI\ column density map overlaid on
    the DSS image.]{ESO274-G001. \HI\ column density map overlaid on
    the DSS image. Contours are plotted in
    $(3,5,10,25,50) \times \sigma$, where the rms noise $\sigma =
    2.02$ \mjybeam\ or $2.30\times 10^{19}$ \atomscmsq. The FWHM
    synthesised beam has dimensions $9.8\arcsec\times9.8\arcsec$, and is
    displayed as a cross-hatched symbol in the lower left.}
  \label{fig:ch2-eso274-g001-m0}
\end{figure}

\addtocounter{figure}{-5}
\begin{figure}[t]
\centering
\includegraphics[width=9cm]{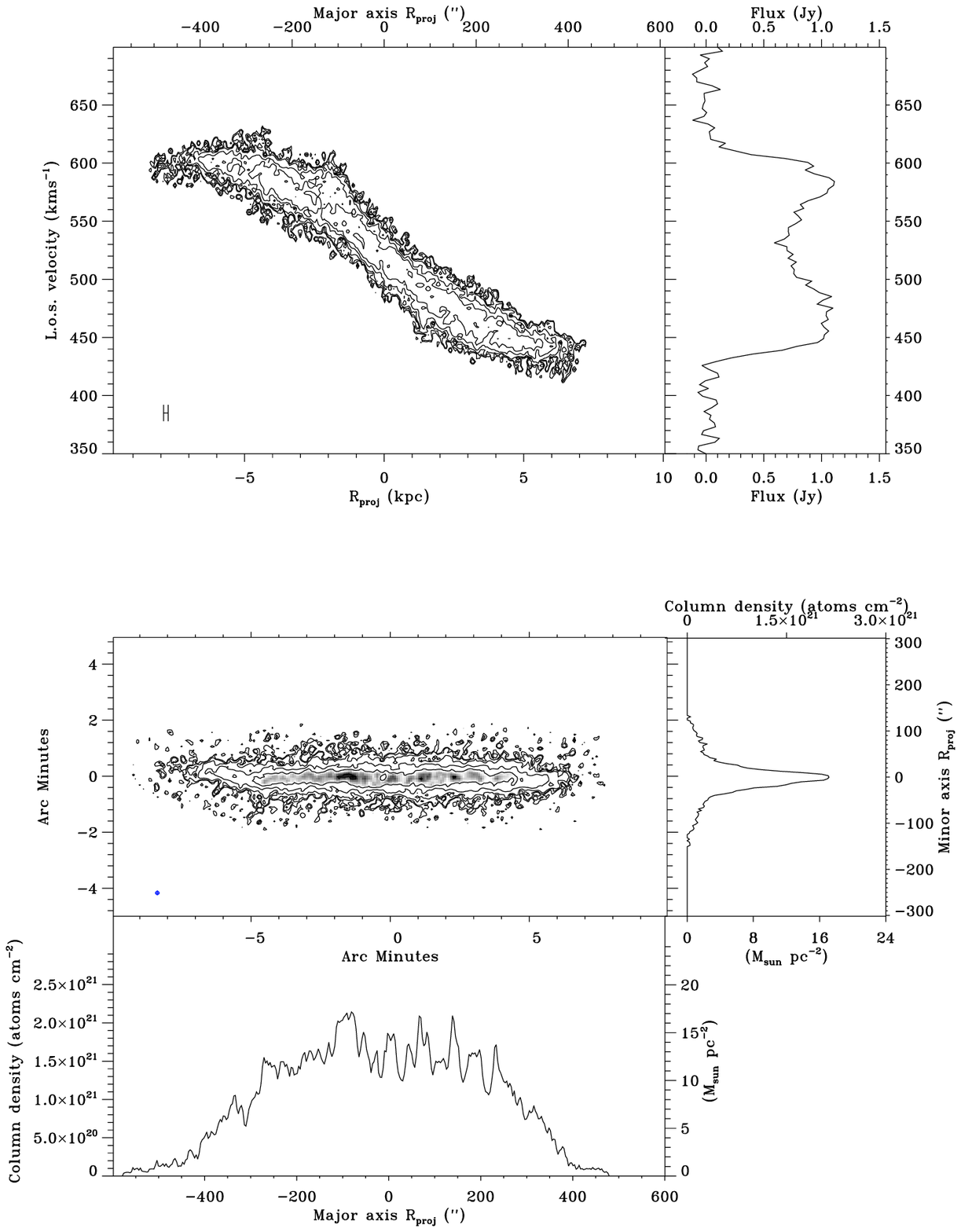}
  \caption[ESO274-G001: XV map, rotated \HI\ column density
  map, integrated \HI\ spectrum and major and minor axis \HI\
  profiles]{ESO274-G001. Top left: XV map. XV contours are
    $(3,5,10,25,50,100)\times \sigma$, where the rms noise
    $\sigma = 2.24$ \mjybeam\ or $2.56\times 10^{19}$ \atomscmsq. The
    half power beam extent over the major axis is shown in the lower
    left corner. Top right: Integrated spectrum. Middle
    left: \HI\ column density map rotated with the galaxy major axis
    aligned with the X axis. Column density map contours are
    $(3,5,10,25,50,100) \times \sigma$, where the rms noise $\sigma
    = 1.70$ \mjybeam\ or $1.93\times 10^{19}$ \atomscmsq. The FWHM
    synthesised beam has dimensions $9.8\arcsec\times9.8\arcsec$, and is
    displayed as a cross-hatched symbol in the lower left.
    Middle right: Minor axis profile. Bottom left: Major axis
    profile.}
  \label{fig:ch2-eso274-g001-combi-plot}
\end{figure}

\addtocounter{figure}{-4}
\begin{figure}[t]
\sidecaption
\includegraphics[width=9cm]{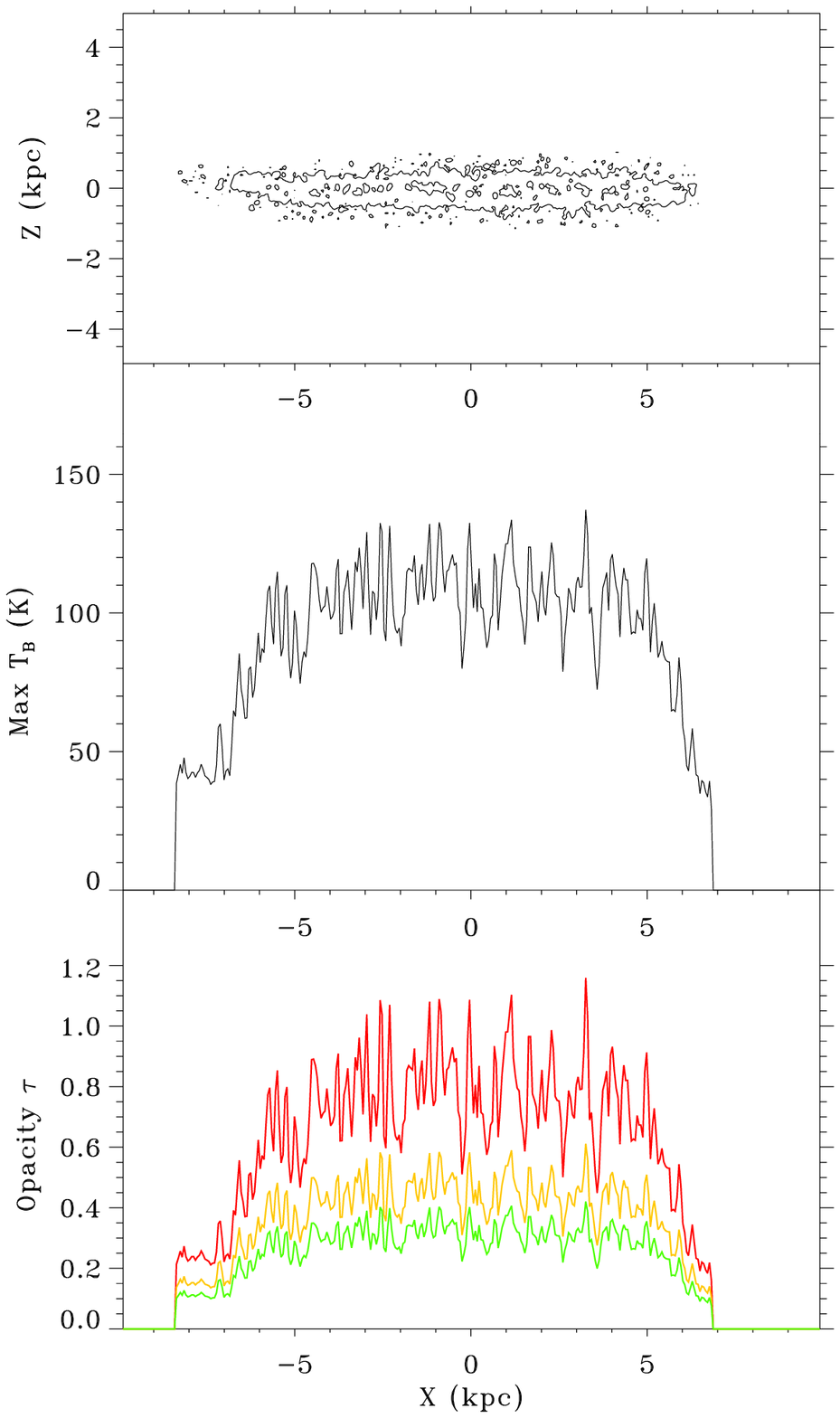}
  \caption[ESO274-G001: \HI\ peak brightness temperature map,
  maximum brightness temperature profile along the major axis, and
  maximum inferred \HI\ opacity along the major axis]{ESO274-G001.
    Top : Peak brightness temperature map. Contours
    are plotted in $(40,100) \times \sigma$, where the rms noise
    is $9.4$ K. The FWHM synthesised beam has dimensions
    $9.8\arcsec\times9.8\arcsec$.  Middle: Major axis
    peak brightness temperature profile.  Bottom: Inferred \HI\ opacity
    calculated assuming constant \HI\ spin temperatures. 
    The resulting maximum opacities along each line of
    sight column through the galaxy disk are plotted 
    for $T_{spin}$ = 200, 300 and  400 K (bottom to top).}
  \label{fig:ch2-eso274-g001-peaktemp-plot}
\end{figure}
\end{subfigures}

\section{Results for individual galaxies}
\label{sec:ch2-results}

In this section we present our data for the galaxies in the sample.
For each galaxy we present the following figures. Part {\em a} shows
an \HI\ column density map overlaid on an optical image from the
Digitized Sky Survey (DSS)\footnote{The Digitized Sky Surveys were
produced at the Space Telescope Science Institute under U.S.
Government grant NAG W-2166. The images of these surveys are based
on photographic data obtained using the Oschin Schmidt Telescope
on Palomar Mountain and the UK Schmidt Telescope. The plates were
processed into the present compressed digital form with the permission
of these institutions.}. Part {\em b} shows the total intensity
position-velocity map (XV map) together with the integrated spectrum,
and the \HI\ column density map aligned with the major axis of the
galaxy and the corresponding major and minor axis surface density
profiles.  Part {\em c} shows the peak brightness temperature map
and the corresponding observed major axis profile and compares this
with opacity profiles calculated with some assumed \HI\ spin
temperatures. The individual channel maps were presented
by \citet{jobrien2007} and are included in further parts {\em d, e} etc.
(depending on how many figures are needed)  as online-only material
at the end of this paper.

\subsection{ESO074-G015 (IC5052)}

ESO074-G015 is a SBd galaxy with a substantial star formation region
extending above and below the galactic plane on the west side of the
galactic centre. However despite this, the \HI\ column density
distribution (shown in Fig.~\ref{fig:ch2-eso074-g015-combi-plot}) is
surprisingly symmetric. Although there is extensive high latitude \HI\
gas extending to $3.7$ kpc above and $4.2$ kpc below the galactic equator,
the distribution of elevated gas appears to be similar on both sides
of the galactic centre. Similarly the XV diagram is also fairly 
symmetric. There is a slight decrease in flux along the line-of-nodes
of the west side of the galaxy possibly indicating gas depletion due 
to a star burst. However
there is no apparent localised region of high velocity dispersion gas
that could be associated with the star burst.

The \HI\ distribution extends to a maximum radial extent of
$281$\arcsec\ or $9.1$ kpc (for the adopted distance of $6.7$ kpc). The
scalelength of the projected distribution $h_X$ is $\approx120$
\arcsec\ or $3.9$ kpc, with the edge of the \HI\ disk at $2.3h_X$. The
exponential
scaleheight of the projected minor axis distribution $h_z$ is $20.5$
\arcsec\ or $667$ pc. The high latitude \HI\ clouds extend to a
height of $3.9$ kpc or $6$ scaleheights away from the equatorial plane.

The galaxy is well-resolved by the synthesised beam along the major
axis, extending to $31$ FWHM beamwidths on each side of the galactic
centre, where each FWHM beamwidth is $292$ pc. In the vertical
direction the measured scaleheight of the projected distribution is
approximately $4 \times$ HWHM beamwidth, indicating that the vertical
\HI\ structure is well-resolved by the synthesised beam. Indeed the
high latitude \HI\ gas extends to $13 \times$ HWHM beamwidth. Vertical
flaring of the gas distribution can be noted in the extreme velocity
channels $499, 509, 657, 667, 677$ \kms\  in the outer part of the
galaxy where the rotation curve inferred from the XV map is roughly
flat.

The \HI\ column density distribution is dominated by two bright spots
at a projected radius of $\sim$1 kpc from the centre. The brightest
gas emission in the cube with brightness temperatures of approximately
$140$ K also correspond to this position. The \HI\ disk also contains
two outer bright points (although less defined on the western side) at
a radius of $\sim$5-6 kpc with a equatorial brightness
distribution similar to that of the barred galaxy UGC7321, which also
displays the characteristic ``figure-8'' pattern of a bar in its
XV distribution.  Although ESO074-G015 is classified as a barred
SBd galaxy in NED, it does not display a ``figure-8'' pattern in
the XV map. ESO074-G015 contains only $1/3$ of the \HI\ mass of
UGC7321, and has a lower maximum rotation speed of $90.1$ \kms,
relative to $112.1$ \kms\ for UGC7321. It is possible that the lower angular
momentum and overall mass of ESO074-G015 restricts gas flow in
the orbital patterns of a barred potential.

\subsection{ESO109-G021 (IC5249)}

ESO109-G021 is a superthin Sd galaxy with B-band stellar
scalelength-to-scaleheight ratio of $11$ \citep{vdkjvkf2001}. The
rotation curve was initially found to be in solid body rotation by
\citet{abeetal1999}, however \citet{vdkjvkf2001} pointed out such a
rotation curve was inconsistent with the double-horned \HI\ spectrum
(see Fig.~\ref{fig:ch2-eso109-g021-combi-plot}). By modelling the XV
diagram \citet{vdkjvkf2001} show that ESO109-G21 is differentially
rotating.

To improve signal-to-noise and recovery of image structure we obtained
two additional $12$-hour synthesis observations at the ATCA and added
these to the earlier observation by Carignan which was used in the
rotation curve analysis by \citet{abeetal1999} and \citet{vdkjvkf2001}.  
The distance of ESO109-G21 is about $30$ Mpc distant so the 
\HI\ beam is relatively large, $1179$ pc. The \HI\ column density map in
Fig.~\ref{fig:ch2-eso109-g021-combi-plot} shows the galaxy to be
well-resolved in the radial direction, while perpendicular to the
plane the resolution is just sufficient to detect flaring in the
extreme velocity channel maps. Comparison of the XV map in
Fig.~\ref{fig:ch2-eso109-g021-combi-plot} with that measured by
\citet{vdkjvkf2001} shows significantly improved structure recovery.
However the peak signal-to-noise of the \HI\ channel maps is $10.1$,
which will make measurement of the \HI\ flaring quite difficult.

\subsection{ESO115-G021}

This Scd galaxy extends to a maximum radial extent of $445$\arcsec\ or
$9.7$ kpc (assuming a Hubble flow distance of $4.5$ kpc inferred from
the $v_{Gal}$). The scalelength of the projected distribution, $h_X$,
is $209$\arcsec\ or $4.6$ kpc, with the edge of the \HI\ disk at
$2.1h_X$. This \HI\ column density map
(Fig.~\ref{fig:ch2-eso115-g021-combi-plot}) displays a very
symmetric disk which appears to be relatively quiescent with no
obvious \HI\ extensions above and below the inner disk. The vertical
structure is well-resolved, extending to a height of $1.7$
kpc, equivalent to 8.5 FWHM beamwidths or $2.9 h_z$, where $h_z$ is
the projected scaleheight of $26.7$ arcsec or $583$ pc . The flaring
of the vertical \HI\ distribution can clearly be seen in channel
maps at $459,
469, 479, 548, 558, 568$ \kms, particularily on the eastern
side where the flatter domain of the rotation curve is more extended
(as inferred from the XV map shown in
Fig.~\ref{fig:ch2-eso115-g021-combi-plot}).

The XV diagram does not display barred gas kinematics; however
a small galaxy like ESO115-G021 with a rotation speed of $62$ \kms\ is
unlikely to show the strong orbital patterns of a gas
bar. The \HI\ column density distribution is dominated by a plateau
within a radius of $5$ kpc, except for a central bright peak with a
peak brightness of $\sim$150 K in the central $1$ kpc.

\begin{subfigures}
\begin{figure}[t]
\centering
\includegraphics[width=9cm,bbllx=65,bblly=230,bburx=435,
        bbury=575,clip=]{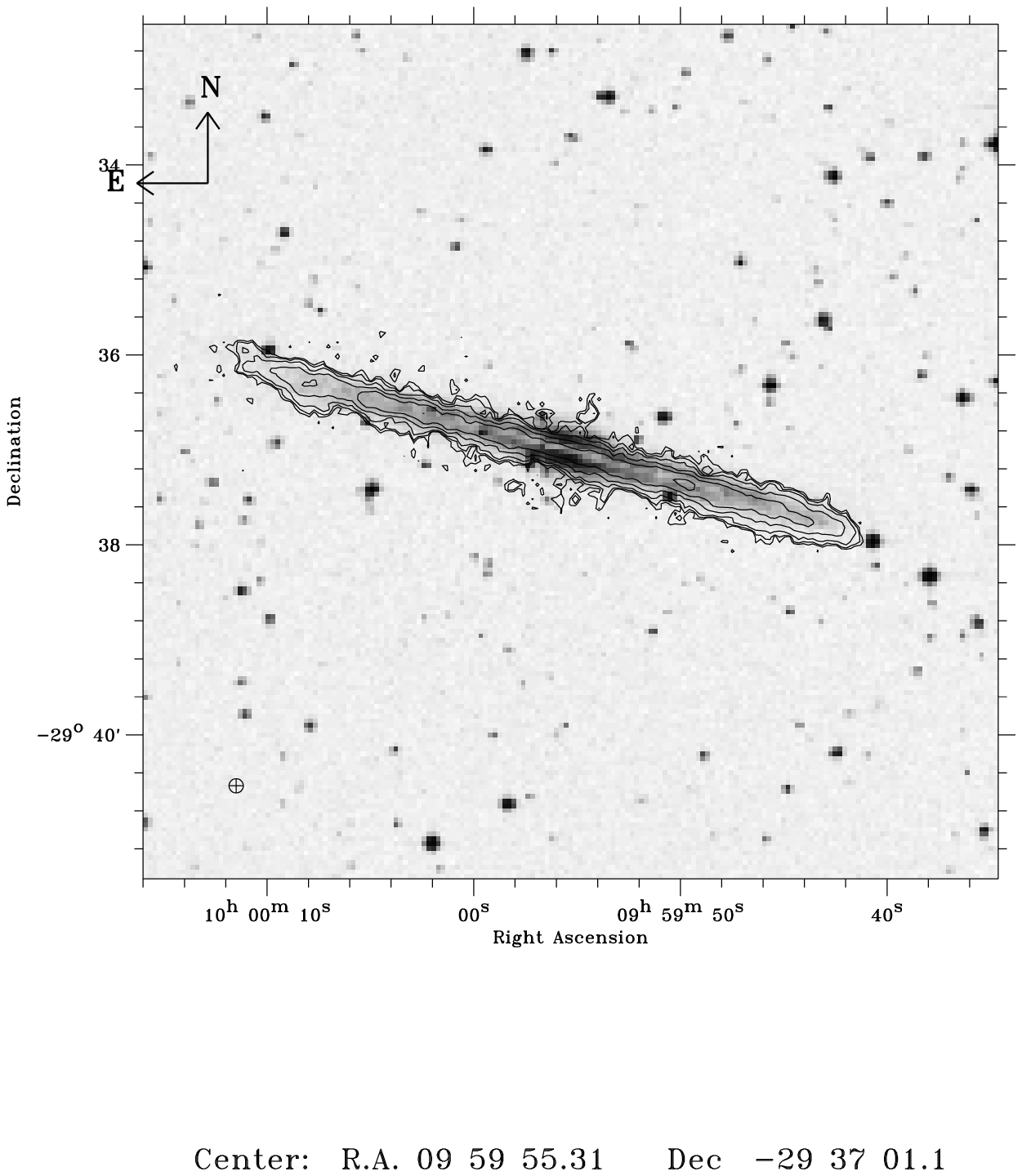}
  \caption[ESO435-G025 (IC2531): \HI\ column density map overlaid on
    the DSS image.]{ESO435-G025 (IC2531). \HI\ column density map overlaid on
    the DSS image. Contours are plotted in
    $(3,5,10,25,50,100) \times \sigma$, where the rms noise $\sigma =
    1.73$ \mjybeam or $2.35\times 10^{19}$ \atomscmsq. The FWHM
    synthesised beam has dimensions $9.0\arcsec\times9.0\arcsec$, and is
    displayed as a cross-hatched symbol in the lower left.}
  \label{fig:ch2-eso435-g025-m0}
\end{figure}

\addtocounter{figure}{-6}
\begin{figure}[t]
\centering
\includegraphics[width=9cm]{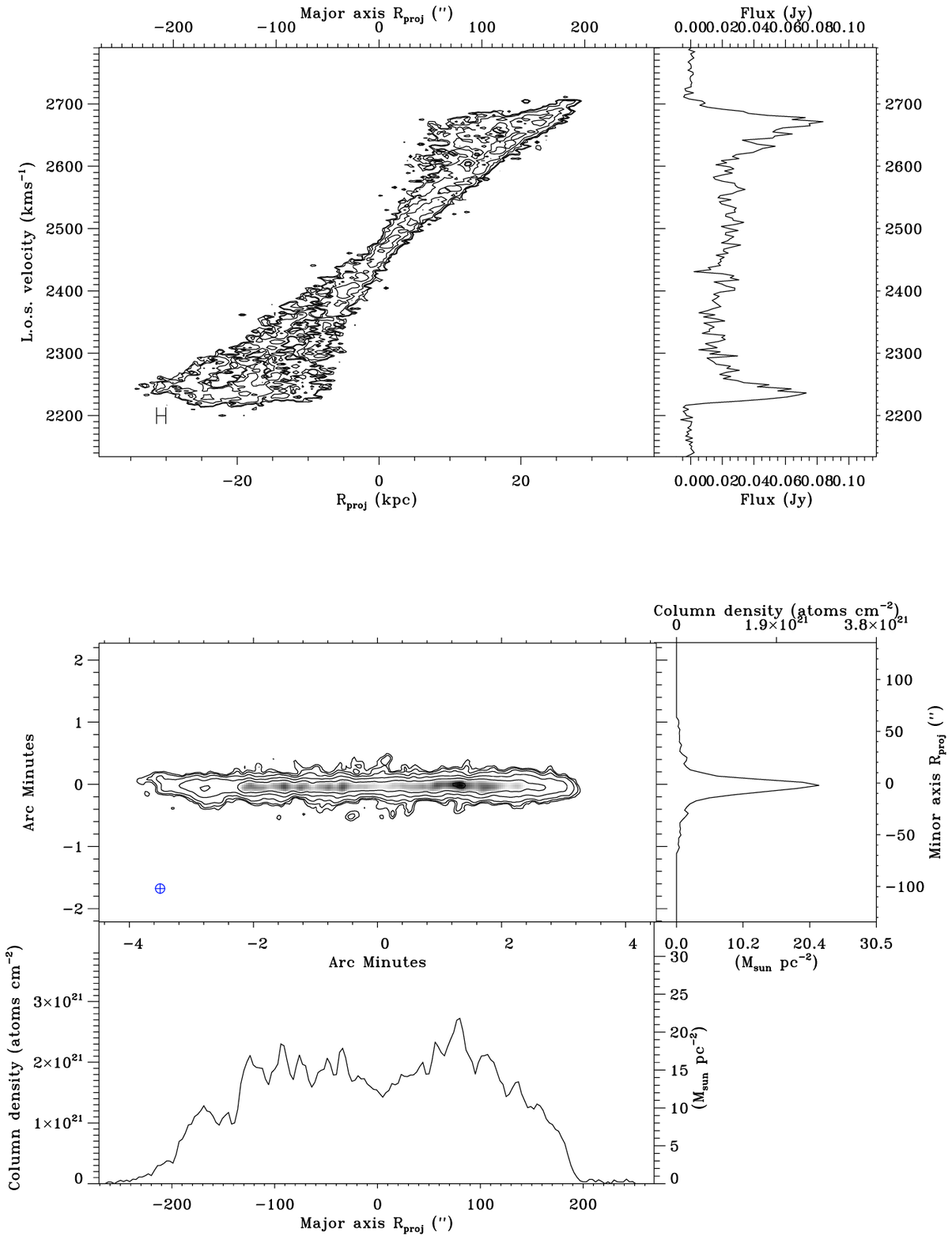}
  \caption[ESO435-G025 (IC2531): XV map, rotated \HI\ column density
  map, integrated \HI\ spectrum and major and minor axis \HI\
  profiles]{ESO435-G025 (IC2531). Top left: XV map. XV contours are
    $(3,5,10,20,30)\times \sigma$, where the rms noise
    $\sigma = 1.76$ \mjybeam\ or $2.40\times 10^{19}$ \atomscmsq. The
    half power beam extent over the major axis is shown in the lower
    left corner. Top right: Integrated spectrum. Middle
    left: \HI\ column density map rotated with the galaxy major axis
    aligned with the X axis. Column density map contours are
    $(3,5,10,25,50,100) \times \sigma$, where the rms noise $\sigma
    = 1.68$ \mjybeam\ or $2.28\times 10^{19}$ \atomscmsq. The FWHM
    synthesised beam has dimensions $9.0\arcsec\times9.0\arcsec$, and is
    displayed as a cross-hatched symbol in the lower left.
    Middle right: Minor axis profile. Bottom left: Major axis
    profile.}
  \label{fig:ch2-eso435-g025-combi-plot}
\end{figure}

\addtocounter{figure}{-5}
\begin{figure}[t]
\sidecaption
\includegraphics[width=9cm]{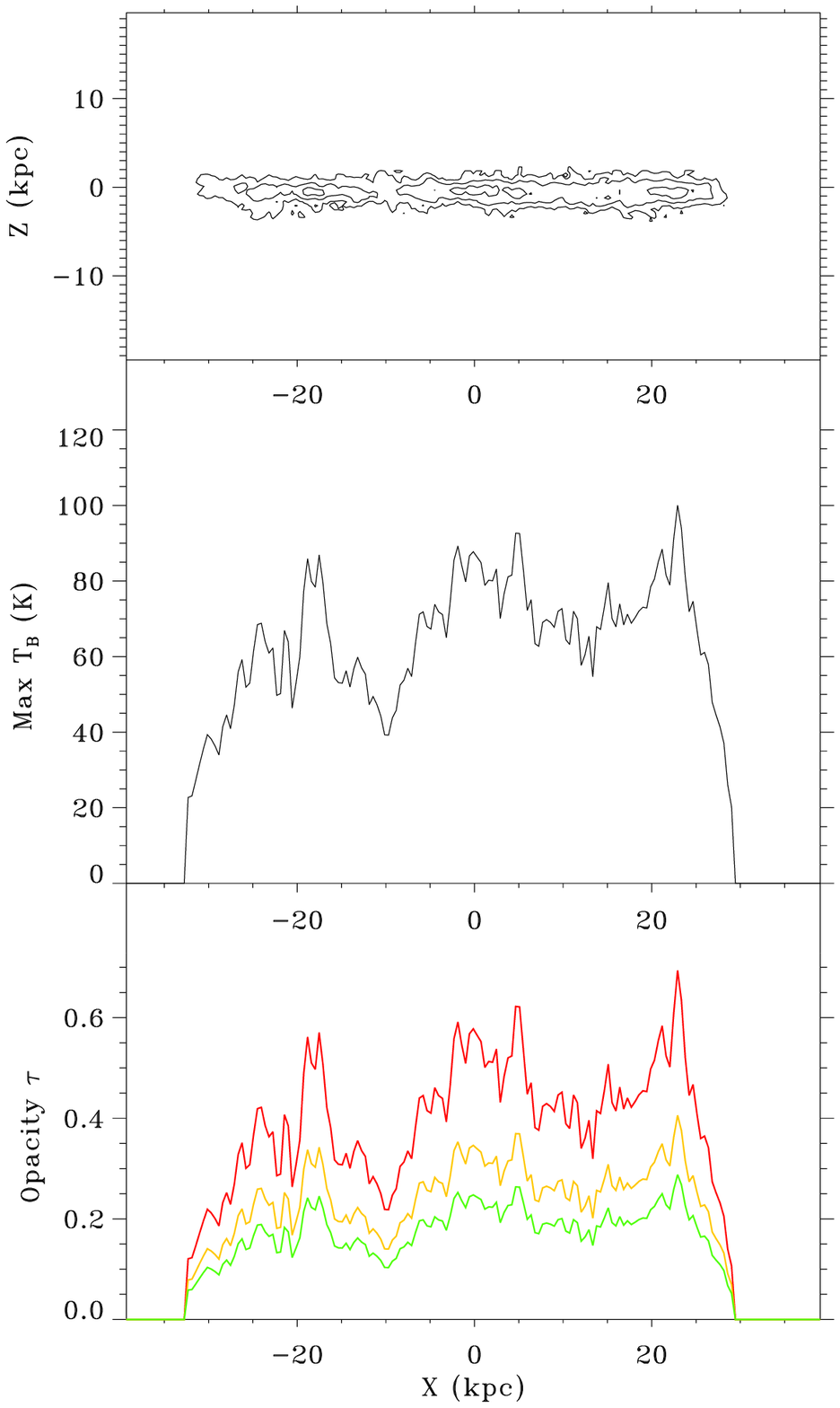}
  \caption[ESO435-G025 (IC2531): \HI\ peak brightness temperature map,
  maximum brightness temperature profile along the major axis, and
  maximum inferred \HI\ opacity along the major axis]{ESO435-G025 (IC2531).
    Top : Peak brightness temperature map. Contours
    are plotted in $(27,46,73) \times \sigma$, where the rms noise
    is $9.2$ K. The FWHM synthesised beam has dimensions
    $9.0\arcsec\times9.0\arcsec$.  Middle: Major axis
    peak brightness temperature profile.  Bottom: Inferred \HI\ opacity
    calculated assuming constant \HI\ spin temperatures. 
    The resulting maximum opacities along each line of
    sight column through the galaxy disk are plotted 
    for $T_{spin}$ = 200, 300 and  400 K (bottom to top).}
  \label{fig:ch2-eso435-g025-peaktemp-plot}
\end{figure}
\end{subfigures}

\begin{subfigures}
\begin{figure}[t]
\centering
\includegraphics[width=9cm, bbllx=42,bblly=195,bburx=462,
        bbury=593,clip=]{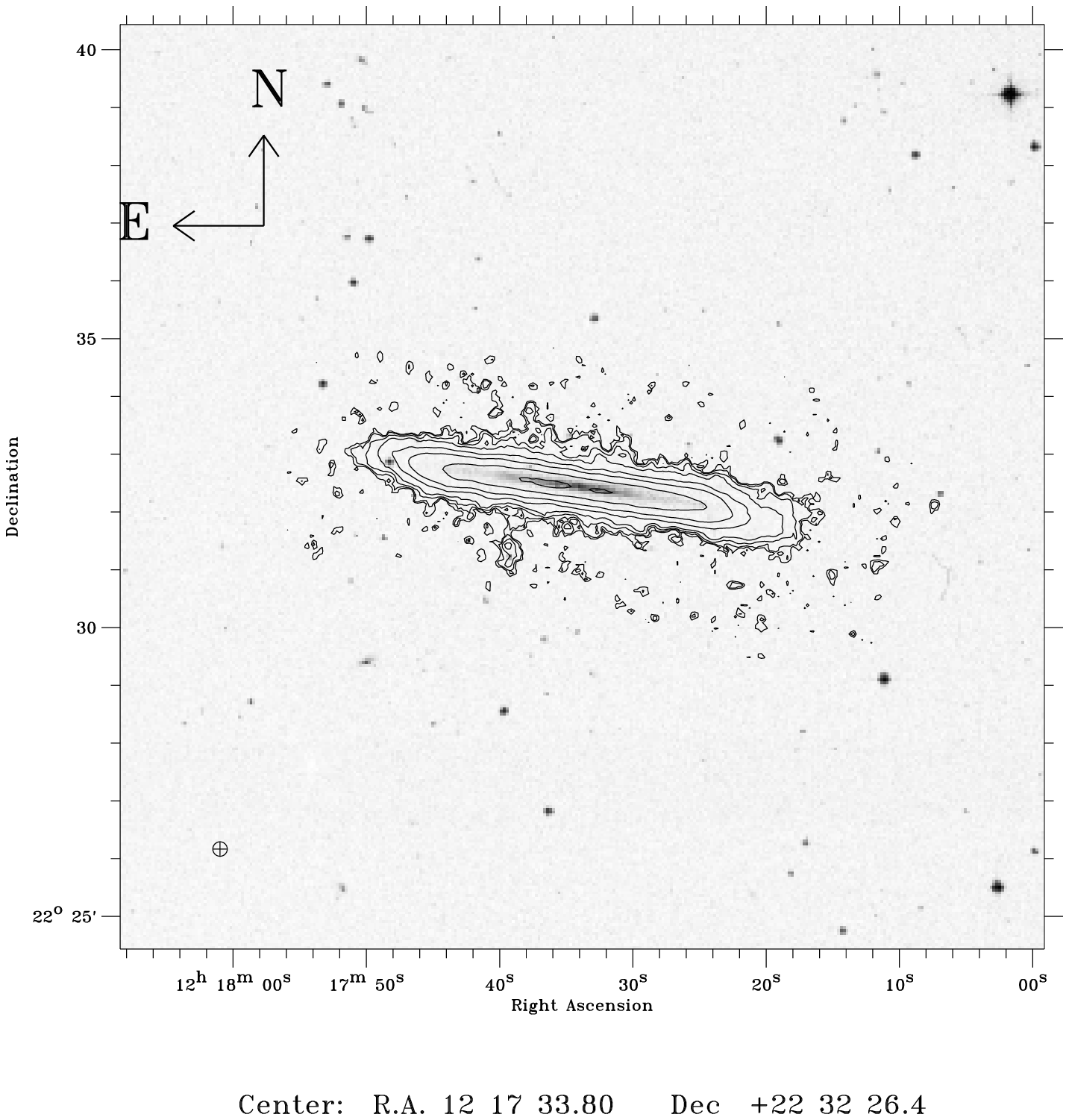}
  \caption[UGC7321: \HI\ column density map overlaid on
    the DSS image.]{UGC7321. \HI\ column density map overlaid on
    the DSS image. Contours are plotted in
    $(3,5,10,25,50,100,250,500) \times \sigma$, where the rms noise $\sigma =$
    0.524 \mjybeam\ or 2.57$\times 10^{18}$ \atomscmsq. The FWHM
    synthesised beam has dimensions $15.0\arcsec\times 15.0\arcsec$, and is
    displayed as a cross-hatched symbol in the lower left.}
  \label{fig:ch2-ugc7321-m0}
\end{figure}

\addtocounter{figure}{-7}
\begin{figure}[t]
\centering
\includegraphics[width=9cm]{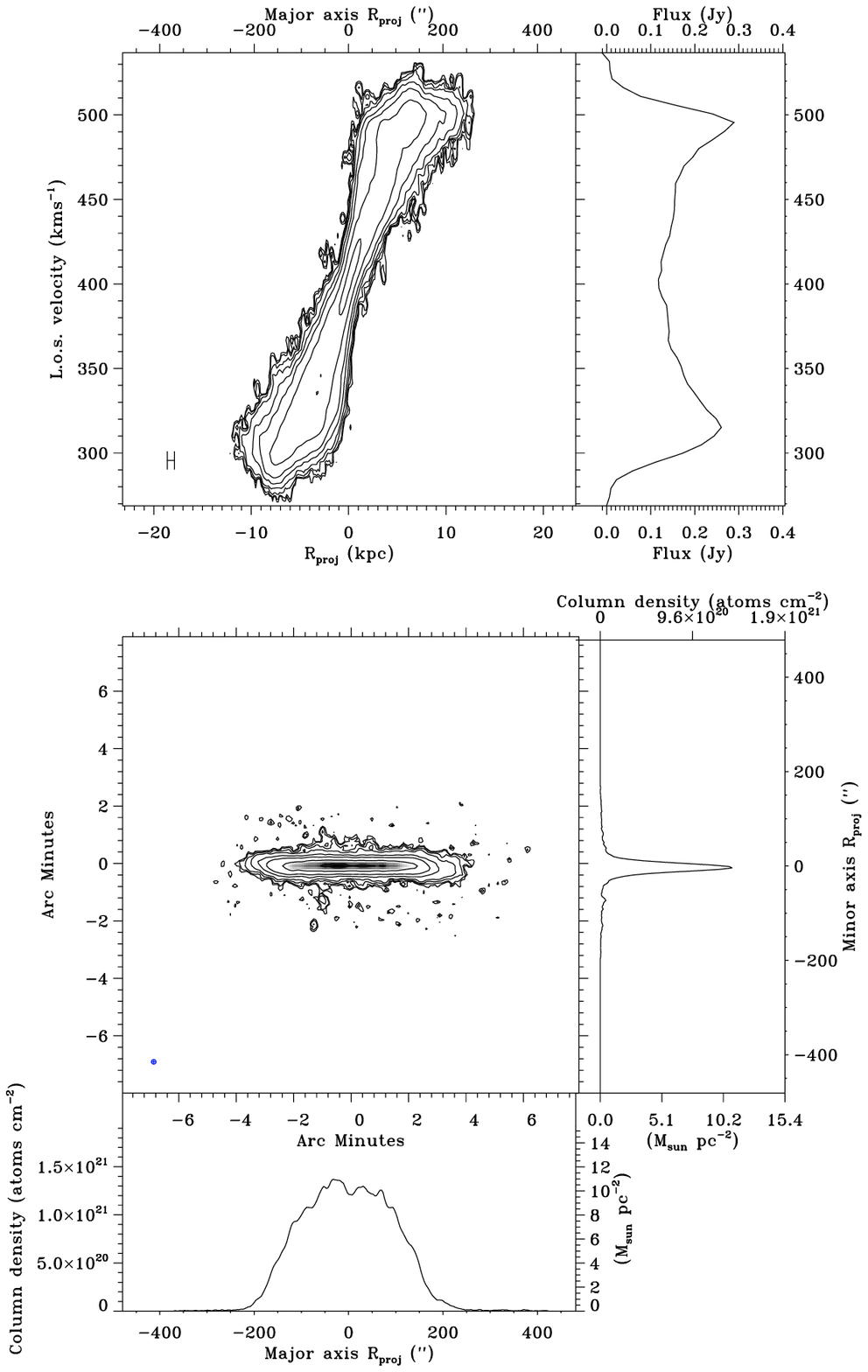}
  \caption[UGC7321: XV map, rotated \HI\ column density
  map, integrated \HI\ spectrum and major and minor axis \HI\
  profiles]{UGC7321. Top left: XV map. XV contours are
    $(3,5,10,20,50,100,250,500)\times \sigma$, where the rms noise
    $\sigma = 0.524$ \mjybeam\ or $2.57\times 10^{18}$ \atomscmsq. The
    half power beam extent over the major axis is shown in the lower
    left corner. Top right: Integrated spectrum. Middle
    left: \HI\ column density map rotated with the galaxy major axis
    aligned with the X axis. Column density map contours are
    $(3,5,10,20,50,100,250,500) \times \sigma$, where the rms noise $\sigma
    = 0.508$ \mjybeam\ or $2.49\times 10^{18}$ \atomscmsq. The FWHM
    synthesised beam has dimensions $15\arcsec\times 15\arcsec$, and is
    displayed as a cross-hatched symbol in the lower left.
    Middle right: Minor axis profile. Bottom left: Major axis
    profile.}
  \label{fig:ch2-ugc7321-combi-plot}
\end{figure}

\addtocounter{figure}{-6}
\begin{figure}[t]
\sidecaption
\includegraphics[width=9cm]{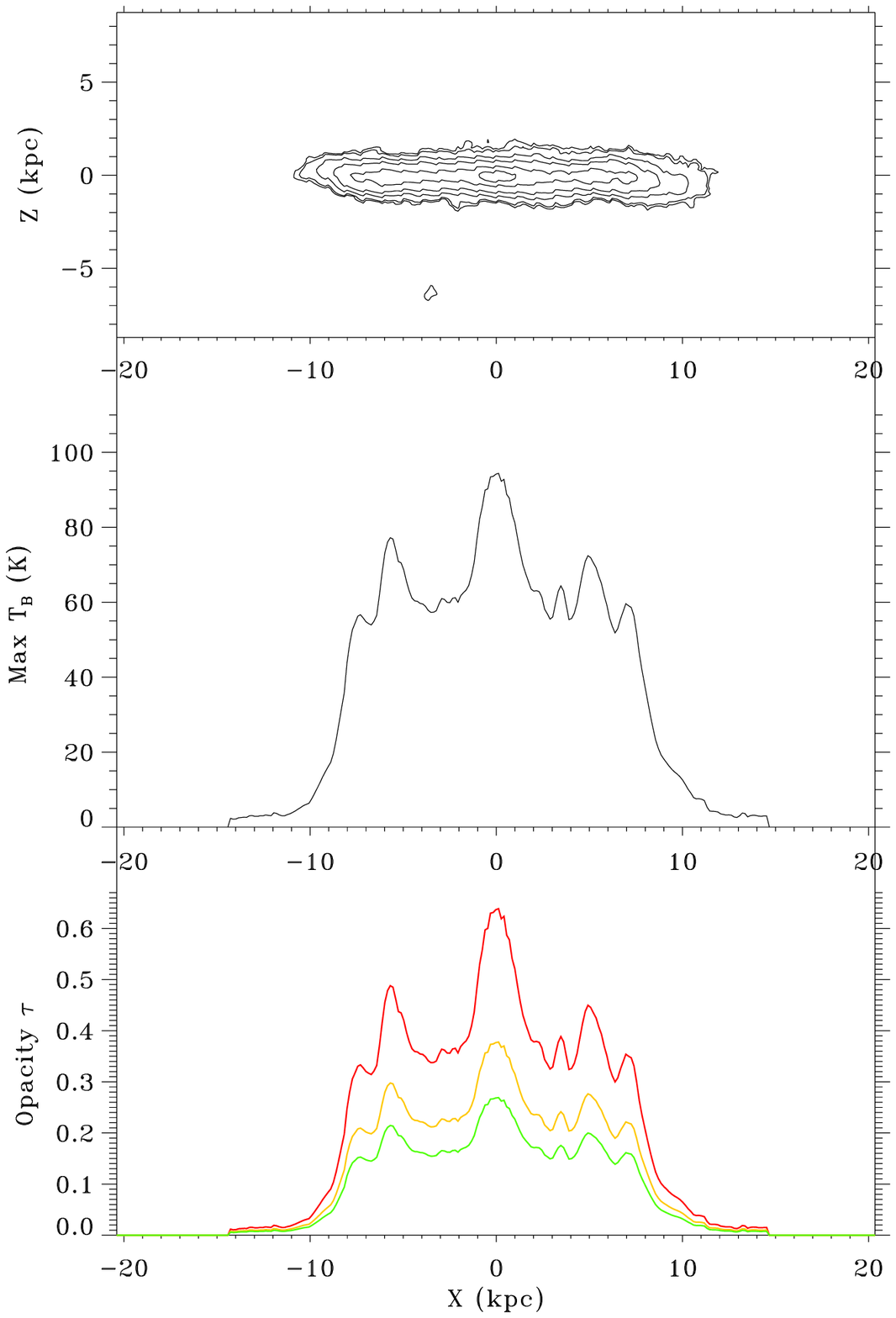}
  \caption[UGC7321: \HI\ peak brightness temperature map,
  maximum brightness temperature profile along the major axis, and
  maximum inferred \HI\ opacity along the major axis]{UGC7321.
    Top : Peak brightness temperature map. Contours
    are plotted in $(4,5,10,20,50,80) \times \sigma$, where the rms noise
    is $1.0$ K. The FWHM synthesised beam has dimensions
    $15\arcsec\times 15\arcsec$.  Middle: Major axis
    peak brightness temperature profile.  Bottom: Inferred \HI\ opacity
   calculated assuming constant \HI\ spin temperatures of $200$, $300$
    and $400$ K. The resulting maximum opacities along each line of
    sight column through the galaxy disk are plotted 
    for $T_{spin}$ = 200, 300, and  400 K (bottom to top).}
  \label{fig:ch2-ugc7321-peaktemp-plot}
\end{figure}
\end{subfigures}

\subsection{ESO138-G014}

ESO138-G14 is a larger Sd galaxy, with a maximum rotation speed of
$120.4$ \kms\ and a \HI\ disk extending to $23$ kpc. The \HI\ column
density map in Fig.~\ref{fig:ch2-eso138-g014-m0} shows the \HI\ disk
is well-resolved by the \HI\ synthesised beam of $10.7$\arcsec\ or
$960$ pc. The XV diagram in
Fig.~\ref{fig:ch2-eso138-g014-combi-plot} shows depressions in the
\HI\ flux at radii of $\sim$10 kpc, with the indication of a
``figure-8'' structure in the gas kinematics. This suggests that the
galaxy is actually barred, as ``figure-8'' structures are typically
seen in gas kinematics of barred edge-on systems \citep{bf1999}.

We were able to observe this galaxy only in the 1.5A and
6A ATCA array configurations. Due to the lack of short spacing
observations we are missing much of the extended structure on large
spatial scales. The channel maps show that the \HI\ disk clearly
flares vertically with radius, but the missing extended structure 
will hamper accurate measurement of the flaring.

\subsection{ESO146-G014} 

This Sd galaxy extends to a maximum radial extent of $118$\arcsec\
or $12.3$ kpc (assuming the distance of $21.5$ kpc determined from
its systemic velocity). The galaxy is well-resolved by the synthesised
beam along the major axis, extending to $15$ FWHM beamwidths on
each side of the galactic centre, where each FWHM beamwidth is $793$
pc. In the vertical direction the measured scaleheight of the
projected distribution is $2.6$ times the HWHM beamwidth, indicating
that the galaxy is well-resolved in the vertical direction also.

ESO146-G014 has a rotation speed of $70.2$ \kms, typical of low-mass
Sd galaxies. The rotation of ESO146-G014 appears solid body from
the XV diagram in Fig.~\ref{fig:ch2-eso146-g014-combi-plot},
however \HI\ modelling (see Paper III) shows a differential inner rotation
curve like other small disk galaxies when observed at high resolution
\citep[e.g.][]{swaters1999}. The XV diagram shows no evidence of a
\HI\ bar, which is as expected given the low oxygen abundance of
$6\%$ of the solar value \citep{br1995}.

\subsection{ESO274-G001}

ESO274-G001 is the closest galaxy in our sample, at a distance of
$3.4$ Mpc in the Centaurus A group. Its proximity allows the \HI\ disk
to be examined at $160$ pc resolution with the $9.8$\arcsec\ resolution
of the \HI\ synthesised beam. To image this galaxy we used two partial
observations in 1.5 km ATCA array configurations from the ATCA archive,
and one new 12-hour synthesis observation in the ATCA 6A array
configuration. From the \HI\ total column density map in
Fig.~\ref{fig:ch2-eso274-g001-m0} we can see that the vertical axis
is very well-resolved, although it is clear from the diffuse structure
that the significant extended \HI\ emission is not observed due to the lack
of observations in short array configurations. Comparison of the \HI\
flux integral with that measured from the \HI\ Parkes All-Sky Survey
(HIPASS) data shows that about 30\% of the \HI\ flux is not
measured.

From the XV map in Fig.~\ref{fig:ch2-eso274-g001-combi-plot} we see
that the rotation curve of this small Sd galaxy rises steeply to a radius of 
$1-2$ kpc and then flattens on one side, but continues to rise with a
shallower slope on the other side. In these outer velocity
channel maps, where the \HI\ spans a larger range along major axis, it
is possible to see clear flaring of the gas thickness with radius.
However, the galaxy is lopsided kinematically, and  also slightly 
lopsided in its \HI\ distribution with an indication of a warp on the NE side.

\subsection{ESO435-G025 (IC2531)} 

ESO435-G25 is the only large spiral galaxy in our sample. It is an Sb
galaxy with a bright peanut-shaped bulge shown to be a bar from the
``figure-8'' signature in the optical emission line kinematics
\citep{bf1999}. The line splitting faintly seen as a figure-8
signature in the \HI\ XV diagram in
Fig.~\ref{fig:ch2-eso435-g025-combi-plot} was first detected by
\citet{bf1997}. We use the same ATCA \HI\ observations in our study.

The \HI\ column density map shows the galaxy extends to $\sim$25
beamwidths on either side of the galactic centre, and three beamwidths
on either side of the \HI\ midplane. Above the galactic centre, the \HI\
column density map in Fig.~\ref{fig:ch2-eso435-g025-m0}
shows \HI\ filaments extending to $\sim$3 kpc above and below the
midplane. Like ESO109-G021, the other distant galaxy, the peak
signal-to-noise is quite low at $10.9$.  Although the \HI\ disk appears
to be flared in the \HI\ column density map
(Fig.~\ref{fig:ch2-eso435-g025-m0}), the flaring is not visibly
obvious in the \HI\ channel maps due to the low
signal-to-noise.

\subsection{UGC7321}

This Sd galaxy extends to a maximum radial extent of $288$\arcsec\ or
$14.0$ kpc (assuming the distance of $10$ kpc adopted by
\citet{um2003}). The scalelength of the projected distribution,
$h_X$, is $112$\arcsec\ or $5.4$ kpc, with the edge of the \HI\ disk at
$2.6h_X$. This galaxy exhibits substantial high latitude \HI\
extending up to $144$\arcsec\ or $7.0$ kpc in the inner \HI\ disk. The
scaleheight of the projected minor axis distribution $h_z$ is $11.6$
\arcsec\ or $560$ pc. Thus the high latitude \HI\ clouds extend to a
height of $12$ scaleheights above the equatorial plane.

The galaxy is well-resolved by the synthesis beam along the major
axis, extending to $19$ FWHM beamwidths on each side of the galactic
centre, where each FWHM beamwidth is $727$ pc. In the vertical
direction the measured scaleheight of the projected distribution is
similar to the FWHM beamwidth, suggesting that the low latitude
vertical structure is dominated by the shape of the synthesised beam.
However despite the relatively large synthesised beam, the vertical
flaring is clearly visible in most channel maps due to the 
extended roughly flat
rotation curve which is clearly apparent in the XV map
(Fig.~\ref{fig:ch2-ugc7321-combi-plot}).

The XV diagram (Fig.~\ref{fig:ch2-ugc7321-combi-plot}) also exhibits
the characteristic ``figure-8'' shape, which can be indicative of
barred \HI\ dynamics \citep{km1995,bf1999,ab1999,athanassoula2000}.
Further evidence for a dynamical bar in UGC7321 was later found by
Pohlen who detected the distinctive boxy-bulge shape in the
near-IR distribution \citep{pbld2003}. \citet{pbld2003} measure the
bar length as $112\pm21$\arcsec\ or $5.4\pm1.0$ kpc. This is in
agreement with the bar size inferred from the size of the ``figure-8''
kinematics. It also corresponds to the scalelength of the
projected \HI\ profile. The \HI\ column density has a plateau at
radii within $5$ kpc, declining steeply at radii outside this radius.

The peak brightness temperature rises steeply at radii inside $2$ kpc
to $94$ K. Assuming a \HI\ spin temperature of $200$ or $400$ K, the
corresponding central \HI\ opacity $\tau$ is $0.65$ or $0.27$. The two
outer peak brightness points correspond to radii of $6$ kpc east, and
$5.5$ kpc west, also aligning with the outer Lindblad resonance points
of the stellar bar measured by \citet{pbld2003}.

\section{Discussion \& Summary}
\label{sec:disc-HI}

The resolution of our \HI\ data along the major axis is high, with
the number of independent beamwidths on each side of the galaxy centre
ranging from $15$ to over $50$ in our galaxies. In the vertical
direction all galaxies are spanned by at least $5$ beamwidths.
Unfortunately three of the galaxies in the sample suffer from
incomplete imaging, resulting in missing information about the
extended spatial structure due to the lack of observations along short
baselines. For two of these galaxies ESO074-G015 and ESO274-G001, the
images still contain substantial information due to the small linear
size of the beam, $290$ pc and $160$ pc, respectively. For these 
galaxies the iterative
\HI\ modelling methods used to measure the deprojected \HI\ density
distribution and kinematics (Paper II) should provide good measurements of 
the flaring.  But the \HI\ images of ESO138-G014 indicate missing 
extended \HI\ emission and low spatial resolution which will probably 
prevent reliable measurement of the flaring in this galaxy. All the 
other galaxies in our sample are promising candidates for accurate
measurement of the \HI\ kinematics and vertical flaring.

The \HI\ column density distribution of the galaxies in our sample
varies quite substantially. The four galaxies with maximum rotation
speeds $\gesim 100$ \kms\ all have \HI\ disks that extend to greater than
$5$ kpc away from the plane. Three of them (ESO435-G25, UGC7321 and
ESO138-G14) also have the ``figure-8'' signature in the XV diagram
suggestive of an edge-on bar in the gas distribution. ESO435-G25
\citep{bf1999} and UGC7321 \citep{pbld2003} both have boxy-peanut
shaped stellar bulges consistent with a bar seen edge-on. Combined
with the figure-8 signature in the \HI\ gas structure this is strong
evidence for a bar \citep{km1995,bf1999,ab1999,athanassoula2000}.
E138-G014, IC5249 and IC2531 all have large \HI\ disks with radii
larger than $20$ kpc, while UGC7321 is much smaller with a radius of
only $14.0$ kpc.

The other four galaxies are not much smaller than UGC7321 in radial
extent, but they are considerably smaller in total mass. They also lack
the vertical extensions in the central disk that suggest additional
sources of heat in the disk. This suggests that one of the causes of
extended high latitude \HI\ filaments is heating related to star
formation and other processes associated with the bar.

The peak brightness profiles of each galaxy show that the inferred \HI\
opacity (assuming a constant \HI\ spin temperature) varies
substantially across the major axis of each galaxy. The high spatial
resolution of our images has made it possible to measure high \HI\
brightness temperatures. The maximum brightness temperature in our
galaxies ranges from $94.4$ K for UGC7321 to $168.6$ K for ESO146-G14.
In addition to ESO146-G14, two other galaxies have high \HI\ brightness
temperatures $> 150$ K. Assuming a mean \HI\ spin temperature of $300$
K, the maximum inferred opacity for these \HI\ bright galaxies is
$\sim$0.7. This is comparable to the maximum \HI\ opacity of $0.85$
measured in NGC891 \citep{kvdkdb2004}. Inspection of the peak
brightness as a function of major axis position for these three
galaxies shows that, in each case, the region of increased opacity is
localised spanning a projected radius of $\sim$1-2 kpc. In ESO115-G21,
this increased opacity region occurs at the galactic centre. However
for ESO138-G14 and ESO146-G14, the regions of increased opacity occur in
the outer disk.
Three of our eight galaxies appear to be warped, and one
is lopsided. This is consistent with results from larger samples,
as e.g. mentioned in the review by \citet{sancisi2008}.

\begin{acknowledgements}

JCO thanks E. Athanassoula,
M.  Bureau, R.  Olling, A. Petric and J. van Gorkom for helpful
discussions.  JCO is grateful to B. Koribalski, R. Sault, L.
Staveley-Smith and R.  Wark for help and advice with data reduction and
analysis. We thank the referee, J.M. van der Hulst, for
his careful and  thorough reading of the manuscripts of this series of papers and 
his helpful and constructive remarks and suggestions.

\end{acknowledgements}

\clearpage

\Online

\begin{appendix} 
\section{Data reduction}

\subsection{Calibration}
\label{sec:cal-HI}

Reduction of the \HI\ data was performed with the radio interferometry
data reduction package {\sc miriad}. First, the data was loaded into
{\sc miriad} with the task {\sc atlod}, using the barycentric velocity
reference frame, and specifying the rest frequency of \HI. The
visibilities were then split into separate datasets for each source.
The primary and secondary calibrators for each source were inspected
for the characteristic signs of solar and terrestrial radio frequency
interference (RFI) using the task {\sc uvplt} and {\sc tvflag}.
Visibilities with RFI, Galaxy \HI\ emission, shadowing or during known
telescope system problems were flagged bad.  The primary calibrator
data were flux calibrated using the known flux with the task {\sc
  mfcal}. The secondary calibrator was then used to calibrate the
telescope gains, and if sufficient parallactic angle coverage was
obtained with the secondary calibrator it was also used to calibrate
the bandpass. For observations with low parallactic angle
coverage, such as snapshots, the bandpass solution of the primary
calibrator was used to calculate the bandpass.

The calibrated flux solution of both primary and secondary calibrators
was inspected with {\sc uvplt}. New secondary calibrators were
calibrated during observations, and then inspected with the
task {\sc uvflux} to see if they resembled an unpolarised point
source. If the secondary calibrator was too faint, showed polarisation
or structure in L band then it was replaced by the next nearest
suitable bright calibrator for later calibrator scans throughout the
observing run.  Gains and bandpass solutions were inspected with the
task {\sc gpplt}. The flux density of the secondary calibrator
was then corrected using the primary calibrator flux solution.

The final calibration tables were copied to the target galaxy
visibility dataset. Before applying the calibration, the gains were
averaged over the duration of each secondary calibrator pointing (3
minutes) and gains interpolation was limited to the interval between
secondary calibrator observations (30-45 minutes).

The calibrated target galaxy visibilities were then inspected with the
tasks {\sc tvflag}, {\sc uvplt}, and any data contaminated by
RFI or shadowing was flagged bad with the appropriate flagging
tool ({\sc tvflag}, {\sc uvflag} or {\sc blflag}).

\subsection{Continuum subtraction}
\label{sec:cont-subtr-HI}

To identify the presence of radio continuum sources in the field
of view, a large $30\arcmin$ channel-averaged (ch0) image was made
using the task {\sc invert} extending out to approximately the half
power radius of the primary beam. The positions and flux in radius
were fitted for any continuum sources found in the ch0 map using
the task {\sc cgcurs}. The visibility spectra of each baseline were
then inspected using the task {\sc uvspec}.  Spectral channels free
of \HI\ emission were identified. Observations of nearby galaxies
often showed Galactic \HI\ emission within the observed bandpass.
It was not necessary to flag these channels.  Instead they were
excluded from the list of line-free channels used in the continuum
subtraction.  As the target galaxy \HI\ emission was always separated
from the Galaxy line emission by at least $100-200$ \kms, they were
also not included in the galaxy \HI\ image cubes.

The radio continuum was then removed by performing a low order
polynomial fit to the line-free channels with the task {\sc uvlin}. If
the observation included a radio continuum source with flux
significantly brighter than the channel-averaged \HI\ flux of the
target galaxy, then the phase center was shifted to the position of
the brightest contaminating continuum source during the continuum
subtraction.  The continuum-corrected visibility spectra were checked
with the task {\sc uvspec} and a new large ch0 image was formed to
search for additional fainter residual continuum sources that remained
after the first continuum subtraction. If found, continuum subtraction
was repeated with the phase center shifted to the next brightest
source.  The process was repeated until the ch0 image showed no
significant contaminating continuum sources. For observations that
were taken during daylight hours, the continuum subtraction was performed 
with the phase center shifted to the position of the Sun. This 
procedure could be compromised by continuum emission
from the galaxy itself or background sources and in all cases it worked 
satisfactorily.

For the narrow bandpass VLA observations, it was not
possible to span the full galaxy emission in a single bandpass when
using the high spectral resolution correlator configuration. In such
cases, the galaxy was imaged with two overlapping bandpasses. As these
bandpasses included only a few line-free channels on one side of the
bandpass it was not possible to perform an accurate 1st order
polynomial fit to the line-free channels with {\sc uvlin}. Instead any
bright continuum sources were subtracted from each bandpass with the
task {\sc uvmodel} using a point source model with the fitted
flux and position of the radio source. Any residual continuum emission
was removed with a zero-order fit to the few line-free channels in
each bandpass. This performed a satisfactory continuum subtraction for
the three galaxies observed at the VLA.

\section{Imaging}
\label{sec:imaging-HI}

A preliminary dirty \HI\ data cube and dirty beam map was formed
using the default gridding where each pixel size is $1/3$
of the synthesised beam FWHM along the X
and Y axes of the beam. The accurate beam shape was then measured from
the central peak of the dirty beam by fitting an elliptical Gaussian
to determine the major $\theta_{maj}$ and minor $\theta_{min}$ axes
and position angle of the synthesised beam. The preliminary data cube
was used to determine the size and velocity range of the \HI\ emission.
A new dirty datacube was then made with square pixels each of $1/3$ or
$1/5$ of the synthesised beam FWHM major axis $\theta_{maj}$.

The visibilities were weighted using a robust weighting of 0.4
\citep{briggs1995} to
provide a optimal compromise between spatial resolution and
suppression of sidelobes. A positive robust weighting causes the
visibilities in the
more sparsely-sampled parts of the visibility phase space to be
down-weighted relative to uniform weighting. This reduces the
amplitude of the sidelobes, without significantly increasing the size
of the synthesised beam FWHM. The barycentric velocity reference frame and the
{\it  radio} definition of velocity was used for the velocity axis of the
\HI\ data cube. We chose the radio definition because
it yields a constant velocity increment between channels, which was
useful for later modelling of the \HI\ distribution.\footnote{There are two
common approximations to the relativistic equation describing the
doppler effect -- the ``radio definition'' for which equal increments
in frequency translate to equal increments in velocity,
$$
    v_{\rm radio}=c(1-\frac{\nu}{\nu_0}),
$$
  and the ``optical definition'',
$$
    v_{\rm optical}=c(\frac{\nu_0}{\nu}-1).
$$
The difference in $v$ for fixed $\nu$ is $\sim$1\%\ for 2500 \kms, the
highest radial velocity in our sample.
  }

We used the {\sc miriad} {\sc clean} algorithm to correct for effects
introduced by the sparcity of the visibility sampling. To find the
approximate \HI\ emission region in a each channel, first an initial
{\sc clean} was performed over the central quarter of each channel
map, and a low resolution data cube was made by restoring the clean
components with a Gaussian of twice the full resolution synthesised
beam size. The noise in each plane of the low resolution cube was then
measured, and inspected to find the region of \HI\ emission in each
plane. Following standard practise, a {\it clean} mask was created by
selecting the region in each plane of the low resolution data cube
with \HI\ emission greater than twice the rms noise ($2\sigma$) using
the task {\sc cgcurs}. The noise level of the full resolution
dirty cube was measured and then the areas defined by the {\it clean}
mask were {\it cleaned} down to $0.5\sigma$ using {\sc miriad} {\sc clean}.
We realise that this is not the optimal procedure
to find the low level emission as CLEAN components found in a high
resolution image usually do not represent the low level extended
emission properly. Better practice would have been to make low resolution images
from the continuum subtracted u,v data directly and CLEAN those. In the
case of this study, however, only high resolution images are produced and
used. 

The final full resolution \HI\ data cube was produced using the
task {\sc restor} which produces a residual map for each channel
map by subtracting the transformed {\em clean} components, and then
adds the restored flux
to each residual image by convolving the {\em
  clean} components by the {\em clean} beam. A circular Gaussian was
used for the {\em clean} beam, with the FWHM set by the measured major axis
FWHM of the dirty beam. After restoration the full cube was corrected
for the effect of the antennae primary beam using the task {\sc
  linmos}. The size of the restoring beam is shown in column 2 \& 3 of
Table~\ref{tab:resol-HI}.

As multiple observations in different array configurations were
combined to provide the maximum visibility sampling, resolution and
sensitivity,
some galaxies were observed at slightly different pointing
positions. These observations were imaged using the {\em mosaic} mode
in {\sc invert}, and deconvolved using the task {\sc mosmem}
which performs a maximum entropy deconvolution of data with multiple
pointings.

In the case of VLA data for which multiple bandpasses were needed to span
the galaxy \HI\ emission, the visibility data were split into separate
bandpasses prior to imaging, with the overlap region combined using
the task {\sc uvcat}. Each bandpass was then imaged, cleaned
and restored separately, before concatenating the final subcubes with
the task {\sc imcat}. Although the channel separation at the
joins was not exactly the same as the channel width elsewhere, the
difference was less than $1$ \kms, which is unlikely to effect
subsequent modelling of the \HI\ distribution. The
overlap channels were observed for a longer integration time and
consequently these channel maps have a lower rms noise. However as
channel maps and moment analysis were made using the noise
measured in each channel map, the difference in noise level over the
velocity range of the galaxy is handled appropriately and does not
cause noise contamination in the derived images. The measured noise
levels of the \HI\ data cubes are shown in Table~\ref{tab:noise-HI}.

After forming the full resolution \HI\ data cubes with Miriad,
all subsequent imaging, moment analysis and measurements were
undertaken using the {\em Interactive Data Language} (IDL).

\section{Channel maps}

In \citep{jobrien2007} all the channel maps that contain \HI\
emission have been dispayed as contour diagrams. Interested readers can obtain 
an electronic .pdf file with these maps upon request from P.C. van der Kruit
or K.C. Freeman at the email address given on the first page.

\end{appendix}

\end{document}